\documentclass[12pt]{article}
\usepackage{amsfonts}
\usepackage{amssymb}
\usepackage{amsbsy}
\usepackage{graphics}

\input epsf.sty

\newlength{\TZ}
\newcommand{\BEQ}{\begin{equation}}     
\newcommand{\BEA}{\begin{eqnarray}}
\newcommand{\EEQ}{\end{equation}}       
\newcommand{\EEA}{\end{eqnarray}}
\def\be{\begin{equation}}
\def\ee{\end{equation}}
\def\ba{\begin{eqnarray}}
\def\ea{\end{eqnarray}}
\newcommand{\eps}{\varepsilon}          
\newcommand{\vph}{\varphi}              
\newcommand{\D}{{\rm d}}                
\newcommand{\II}{{\rm i}}               
\newcommand{\wit}[1]{\widetilde{#1}}    
\newcommand{\lap}[1]{\overline{#1}}     

\renewcommand{\vec}[1]{{\boldsymbol{#1}}} 

\newcommand{\zeile}[1]{\vskip #1 \baselineskip} 

\newcommand{\appsection}[2]{\setcounter{equation}{0} \section*{Appendix #1. #2}
\renewcommand{\theequation}{#1\arabic{equation}}
              \renewcommand{\thesection}{#1} }

\catcode`\@=11
\def\numberbysection{\@addtoreset{equation}{section}
        \def\theequation{\thesection.\arabic{equation}}}
\numberbysection

\begin{document}

\begin{titlepage}

~~~ 

\vskip 1.5 cm
\begin{center}
{\Large \bf The kinetic spherical model in a magnetic field}
\end{center}

\vskip 2.0 cm
\centerline{{\bf Matthias Paessens}$^{a,b}$ and {\bf Malte Henkel}$^{a}$}
\vskip 0.5 cm
\centerline {$^a$Laboratoire de Physique des
Mat\'eriaux,\footnote{Laboratoire associ\'e au CNRS UMR 7556}
Universit\'e Henri Poincar\'e Nancy I,}
\centerline{ B.P. 239,
F -- 54506 Vand{\oe}uvre l\`es Nancy Cedex, France}
\vskip 0.5 cm
\centerline{$^b$Institut f\"ur Festk\"orperforschung (Theorie II), 
Forschungszentrum J\"ulich,
D -- 52425 J\"ulich, Germany}

\begin{abstract}

The long-time kinetics of the spherical model in an external magnetic field 
and below the equilibrium critical temperature is studied. The solution of 
the associated stochastic Langevin equation is reduced exactly to a single 
non-linear Volterra equation. For a sufficiently small
external field, the kinetics of the magnetization-reversal transition from the 
metastable to the ground state is compared to the ageing behaviour of 
coarsening systems quenched into the low-temperature phase. For an oscillating
magnetic field and below the critical temperature, we find evidence for 
the absence of the frequency-dependent dynamic phase transition, which was
observed previously to occur in Ising-like systems. 
\end{abstract}

\zeile{4}
\end{titlepage}

\section{Introduction}

Non-equilibrium critical phenomena are a subject of intense research 
activity. A common way to reach such a situation is through a rapid
change of one of the macroscopic variables which enter into the equation
of state. For definiteness, consider a simple ferromagnet. It may
be brought out of equilibrium, for example, by starting initially
from a fully disordered state and then quench the system rapidly to a
temperature below the system's critical temperature $T_c>0$. The resulting
ageing behaviour has been in the focus of intensive study, see 
\cite{Stru78,Bray94,Cate00,Godr02,Cugl02} for reviews. Another way to
reach a non-equilibrium state is to start from a magnetically ordered
state below $T_c$ and then turn on a magnetic field $H$ oriented antiparallel
with respect to the magnetic order parameter. Then the system will find itself
in a metastable stable and a magnetization reversal transition towards the
stable ground state will take place, see \cite{Seth97,Rikv01} for reviews.

After a quench to below $T_c$, the system undergoes phase-ordering, that
is domains of a time-dependent typical size $L(t)\sim t^{1/z}$ form and
grow, where $z$ is the dynamical exponent. As a consequence, a system
of infinite size slowly evolves towards an equilibrium state, without ever
reaching it. This evolution is more fully revealed through the
study of {\em two-time} quantities, such as the two-time autocorrelation
function $C(t,s)$ and the autoresponse function $R(t,s)$
\BEQ
C(t,s) = \langle \phi(t) \phi(s) \rangle \;\; , \;\;
R(t,s) = \left.\frac{\delta\langle\phi(t)\rangle}{\delta h(s)}\right|_{h=0}
\EEQ
where $\phi$ is the order parameter, $h$ the conjugate magnetic field,
$t$ is called the observation time and $s$ the waiting time. Ageing 
occurs in the regime when $s$ and $\tau=t-s>0$ are simultaneously much larger
than any microscopic time scale $\tau_{\rm micro}$. In many systems, one finds
in the ageing regime a scaling behaviour, see \cite{Cate00,Godr02} 
\BEQ \label{gl:CR}
C(t,s) = s^{-b} f_C(t/s) \;\; , \;\; R(t,s) = s^{-1-a} f_R(t/s)
\EEQ
where $a$ and $b$ are non-equilibrium exponents. For $T<T_c$, $b=0$ while
$a$ depends on whether there are short-ranged or long-ranged correlations in
the equilibrium state. For short-ranged correlations, $a=1/z$, whereas for
long-ranged correlations $a=(d-2+\eta)/z$ \cite{Henk02d}. The scaling functions
behave for large arguments $x=t/s\gg 1$ asymptotically as
\BEQ
f_C(x) \sim x^{-\lambda_C/z} \;\; , \;\; f_R(x) \sim x^{-\lambda_R/z}
\EEQ
where $\lambda_C$ and $\lambda_R$ are the autocorrelation \cite{Fish88,Huse89} 
and autoresponse \cite{Pico02} exponents, respectively. While for a fully
disordered initial state, it is traditionally accepted that 
$\lambda_C=\lambda_R$, for spatially long-ranged initial correlations of
the form $C_{\rm ini}(\vec{r})\sim |\vec{r}|^{-d-\alpha}$ (with $\alpha\leq 0$)
the relation $\lambda_C=\lambda_R+\alpha$ has been conjectured \cite{Pico02}.
Furthermore, the rigorous arguments of \cite{Yeun96} readily yield
$\lambda_C\geq (d+\alpha)/2$. Very recently, different exponents 
$\lambda_C\ne\lambda_R$ have also been found in the random sine-Gordon model 
and in addition $\lambda_C< d/2$ violates the rigorous bound mentioned 
above \cite{Sche03}. 
In addition, and going beyond these traditional scaling arguments, 
it has been proposed recently that the dynamical symmetry group of
ageing systems might include more general transformations than merely
the simple dynamical scaling as expressed by eq.~(\ref{gl:CR}). In particular,
there is evidence that the dynamical group of ageing systems 
includes so-called local scale transformations related to conformal 
transformations in time \cite{Henk02}. 
If that is the case, the form of the scaling function 
\BEQ
f_R(x) = r_0 x^{1+a-\lambda_R/z} (x-1)^{-1-a}
\EEQ
is completely fixed ($r_0$ is a normalization constant) \cite{Henk01,Henk02}.
Going beyond phenomenological tests, at least for the case $z=2$ it can be
shown that, given only the covariance of the response functions under scale
and also Galilei transformations, then a Ward identity guarantees the 
covariance under the full group of local scale transformations \cite{Henk03}. 
Furthermore, the causality condition $R(t,s)=0$ for $t<s$ also follows in a
model-independent way \cite{Henk03}. 
Tests of Galilei invariance require the consideration of 
space-time-dependent response functions, going beyond the autoresponse function
$R(t,s)$ of eq.~(\ref{gl:CR}). Indeed, the phase-ordering kinetics of the 
Glauber-Ising model has recently been shown to be 
Galilei-invariant in the ageing regime \cite{Henk02c}. 

Another central questions in this context is how to characterize
whether/when under the conditions just described the system is in
thermodynamic equilibrium. It is convenient to consider 
the fluctuation-dissipation ratio \cite{Cugl94a,Cugl94b}
\BEQ
X(t,s) = T R(t,s) \left( \frac{\partial C(t,s)}{\partial s}\right)^{-1}
\label{1:eq5}
\EEQ
At equilibrium, the fluctuation-dissipation theorem states that $X(t,s)=1$.
The breaking of the fluctuation-dissipation theorem has been investigated 
intensively both theoretically 
(see e.g. \cite{Cate00,Godr02,Garr01,Pere02,Cugl02}) 
and experimentally \cite{Grig99,Heri02,Bell02}.
 
Here we are interested in the non-equilibrium behaviour associated with the
magnetization-reversal transition from the metastable to the equilibrium
state. This problem has also received intense attention, both
experimentally (e.g. \cite{Wern97,Mang99,Ludw03}) and theoretically, see  
\cite{Bren02,Side99,Korn00,Korn02,Meht01,Chat02,Turk03}. What can be learned 
from the study of two-time quantities about this process ? Additional insight
may be obtained by studying the system's response to a time-dependent, e.g.
oscillating, magnetic field and we shall study whether there exists a dynamic
phase-transition at a finite and non-vanishing value of the period $P$ of the 
field \cite{Tome90,Mend91}. Surprisingly, there is {\em no} such
transition in the spherical model, although it is known to occur e.g. in the
$2D$ Ising model. 

In order to obtain explicit analytical results, we shall study the 
the effects of a magnetic field in the kinetic mean 
spherical model, to be defined precisely in section 2.
This is one of the very few models which can be solved exactly in a great
variety of circumstances and has been studied in detail in the
past, either in the context of continuum field theories
\cite{Bray91,Jans89,Newm90,Kiss93,Coni94,Cala02,Fusc02} or else in the form
of a lattice model 
\cite{Cugl95a,Cugl95b,Zipp00,Godr00b,Cann01,Corb02,Pico02,Pico03}. 
It is known that in $d<4$ dimensions, the spherical model yields results
distinct from mean-field theory and therefore permits the study of fluctuation
effects. In addition, we recall that experimental results of the magnetization
reversal \cite{Wern97,Mang99} are usually described in terms of an 
anisotropic Heisenberg model. Recall that the spherical model shares the
following {\em equilibrium} properties with the O($3$) Heisenberg model and 
which distinguish it from the often-used Ising model:
\begin{itemize}
\item it has a continuous symmetry (O($n$) in the $n\to\infty$ limit).
\item there is no equilibrium phase-transition in $2D$.
\item the equilibrium specific heat exponent $\alpha<0$ in $3D$. 
\end{itemize}
These similarities might suggest that qualitatively the 
kinetics of spherical and the O($3$) Heisenberg models should be closer to 
each other than either is to the kinetics of the Ising model.
Still, the spherical model should be considered a toy model certainly not 
meant to be physically realistic.

This paper is organized as follows. In section 2, the model is defined and 
the exact solution outlined. All physical quantities can be expressed in terms
of the time-dependent solution $g(t)$ of a nonlinear Volterra integral 
equation. In section 3, the solution of this equation and its asymptotics
are discussed. In sections 4 and 5, single- and two-time observables
are calculated for  the full time-range of the 
magnetization-reversal transition for constant magnetic fields and in 
section 6 time-dependent fields are considered. Section 7 presents our 
conclusions. In the appendices, we comment on the numerical techniques and
study the exact long-time behaviour of the Volterra equation.

\section{Model and formalism}

We begin by recalling the definition of the kinetic mean spherical model,
using the formalism as exposed in \cite{Cugl95a,Cugl95b,Godr00b,Pico02}.
We consider a system of time-dependent classical
spin variables $S_{\vec{x}}(t)$ located on the sites $\vec{x}$ of
a $d$-dimensional hypercubic lattice. They may take arbitrary real values
subject only to the mean spherical constraint
\BEQ
\sum_{\vec{x}} \left\langle S_{\vec{x}}(t)^2 \right\rangle = {\cal N}
\label{eq0}
\EEQ
where $\cal N$ is the number of sites of the lattice.
The role of imposing the spherical constraint either microscopically or rather
in the mean (which is the only case where the dynamics can be solved) has been
carefully studied recently \cite{Fusc02}. Provided the infinite-volume limit
is taken {\em before} the long-time limit, either way of treating the spherical 
constraint leads to the same results. 

The spherical model Hamiltonian reads
\BEQ
{\cal H} = - J\sum_{<\vec{x},\vec{y}>}S_{\vec{x}}(t)S_{\vec{y}}(t) 
- \sum_{\vec{x}} H_{\vec{x}}(t) S_{\vec{x}}(t)
\label{eq1}
\EEQ
where $H_{\vec{x}}(t)$ is the space- and time-dependent external magnetic 
field. The first sum extends over 
nearest-neighbour pairs only and the second sum over the entire lattice.  
We choose units such that $J=1$.
The system is supposed to be translation-invariant in all directions.
The kinetics is assumed to be described in terms of a Langevin equation 
\BEQ
\frac{\D S_{\vec{x}}(t)}{\D t} = \sum_{\vec{y}(\vec{x})}S_{\vec{y}}(t)  
-(2d+\mathfrak{z}(t))S_{\vec{x}}(t) + H_{\vec{x}}(t) + \eta_{\vec{x}}(t)
\label{eq2}
\EEQ
where the sum over $\vec{y}$ extends over the nearest neighbours of $\vec{x}$.
The gaussian noise $\eta_{\vec{x}}(t)$ describes that 
the model is in contact with a heat bath. 
It is characterized by a vanishing ensemble-average and the second moment
\BEQ
\langle\eta_{\vec{x}}(t) \eta_{\vec{y}}(t')\rangle =
2 T\,\delta_{\vec{x},\vec{y}}  \delta (t-t')
\label{eq3}
\EEQ
Finally, the function $\mathfrak{z}(t)$ is fixed by the mean spherical 
constraint (\ref{eq0}) and has to be determined.

By a Fourier transformation
\BEQ
\wit{f}(\vec{q}) = \sum_{\vec{r}} f_{\vec{r}} e^{-\II \vec{q}\cdot\vec{r}}
\;\; , \;\;
f_{\vec{r}} = (2\pi)^{-d} \int_{{\cal B}}\!\D\vec{q}\, \wit{f}(\vec{q})
e^{\II \vec{q}\cdot\vec{r}}
\EEQ
where the integral is taken over the first Brillouin zone $\cal B$, the
Fourier-transformed spin variable $\wit{S}(\vec{q},t)$ becomes
\BEQ
\wit{S}(\vec{q},t) = \frac {e^{-\omega(\vec{q})t}}{\sqrt{g(t)}}
\left[\wit{S}(\vec{q},0) +
\int_0^t\!\D t'\:
e^{\omega(\vec{q})t'} \sqrt{g(t')}~ \left[ 
\wit{H}(\vec{q},t')+\wit{\eta}(\vec{q},t')\right] \right]
\label{eq6}
\EEQ
with the dispersion relation
\BEQ
\omega(\vec{q}) = 2\sum_{i=1}^d{\left(1-\cos(q_i)\right)}
\label{eq7}
\EEQ
and we have also defined
\BEQ
g(t)= \exp\left({2\int_0^t\!\D t'\: \mathfrak{z}(t')}\right)
\label{eq5}
\EEQ
Clearly, the time-dependence of $\wit{S}(\vec{q},t)$ and any correlators
will be given in terms of the function $g=g(t)$.

We now derive the expressions for the correlators and response functions for
an arbitrary external field $H_{\vec{x}}(t)$ and general initial conditions. 
Consider the two-time spin-spin correlation function
\BEQ
C_{\vec{x},\vec{y}}(t,s) =C_{\vec{x}-\vec{y}}(t,s) =
\langle S_{\vec{x}}(t) S_{\vec{y}}(s) \rangle
= (2\pi)^{-2d} \int_{{\cal B}^2} \!\D\vec{q}\,  \D\vec{q}'\:
e^{\II(\vec{q}\cdot\vec{x}+\vec{q}'\cdot\vec{y})}\:
\langle\wit{S}(\vec{q},t)\wit{S}(\vec{q}',s)\rangle
\label{eq8}
\EEQ
A straightforward calculation gives 
\BEA
\wit{C}(\vec{q},\vec{q}';t,s) &=& 
\langle \wit{S}(\vec{q},t) \wit{S}(\vec{q}',s) \rangle 
\nonumber \\
&=& \frac{e^{-\omega(\vec{q})t-\omega(\vec{q}')s}}{\sqrt{g(t)g(s)\,}}
\left[
(2\pi)^d \delta^d(\vec{q}+\vec{q}') \left( \wit{C}(\vec{q},t) +  
2T\int_0^t\!\D t'\:  e^{2\omega(\vec{q})t'}g(t') \right) \right.
\nonumber \\
& & + \langle\wit{S}(\vec{q}',0)\rangle \int_{0}^{t}\!\D t'\: 
e^{\omega(\vec{q})t'} \sqrt{g(t')}\: \wit{H}(\vec{q},t') 
+ \langle\wit{S}(\vec{q},0)\rangle \int_{0}^{s}\!\D s'\: e^{\omega(\vec{q'})s'}
\sqrt{g(s')}\: \wit{H}(\vec{q'},s') 
\nonumber \\
& & + \left. 
\int_{0}^{t} \!\D t'\:e^{\omega(\vec{q})t'} \sqrt{g(t')}\: \wit{H}(\vec{q},t')
\int_{0}^{s} \!\D s'\:e^{\omega(\vec{q}')s'}\sqrt{g(s')}\: \wit{H}(\vec{q}',s')
\right]
\label{eq12}
\EEA
where $\wit{C}(\vec{q},t)$ is the single-time correlator. Here the average was
carried out over the noise and the initial conditions 
$S_{\vec{x}}(0)$ such that
\BEQ
\langle\wit{S}(\vec{q},0)\rangle = \sum_{\vec{x}} \langle S_{\vec{x}}(0)\rangle
e^{-\II \vec{q}\cdot\vec{x}} = (2\pi)^d \delta^d(\vec{q})\, S_0
\EEQ
In direct space, the two-time autocorrelator becomes
\BEA
C_{\vec{x},\vec{x}}(t,s) &=& 
(2\pi)^{-2d} \int_{{\cal B}^2} \!\D \vec{q}\, \D \vec{q}'\: 
e^{\II(\vec{q}+\vec{q}')\cdot\vec{x}}\: \wit{C}(\vec{q},\vec{q}';t,s) 
\nonumber \\
&=& \frac{1}{\sqrt{g(t) g(s)\,}}
\left[ A\left(\frac{t+s}{2}\right) + 2T \int_{0}^{s} \!\D u\, 
f\left(\frac{t+s}{2}-u\right) g(u) \right. 
\nonumber \\
& & + S_0 \int_{0}^{t} \!\D t' B_{\vec{x}}(t') \sqrt{g(t')\,} 
    + S_0 \int_{0}^{s} \!\D s' B_{\vec{x}}(s') \sqrt{g(s')\,} 
\nonumber \\
& & \left.+ \int_{0}^{t} \!\D t' B_{\vec{x}}(t') \sqrt{g(t')\,} 
            \int_{0}^{s} \!\D s' B_{\vec{x}}(s') \sqrt{g(s')\,}
\right]
\label{eq13}
\EEA
where we have defined
\BEA
f(t) &=& (2\pi)^{-d} \int_{\cal B} \!\D\vec{q}\: e^{-2\omega(\vec{q})t} 
= \left( e^{-4t} I_0(4t)\right)^d
\nonumber \\
A(t) &=& (2\pi)^{-d} \int_{\cal B} \!\D\vec{q}\: e^{-2\omega(\vec{q})t}\,  
\wit{C}(\vec{q},0) 
\label{2:fAB} \\
B_{\vec{x}}(t) &=& (2\pi)^{-d} \int_{\cal B} \!\D\vec{q}\: 
e^{\omega(\vec{q})t+\II \vec{q}\cdot\vec{x}}\, \wit{H}(\vec{q},t) 
\nonumber 
\EEA
and $I_0$ is a modified Bessel function \cite{Abra65}. We see explicitly
how the initial magnetization $S_0$ and the initial correlator $C_{\vec{x}}(0)$
affect the dynamics of the system. 

It remains to determine the function $g(t)$. 
Because of the spherical constraint (\ref{eq1}) and spatial translation
invariance, the equal-time autocorrelator must satisfy
\BEQ
C_{\vec{0}}(t,t)= \int_{{\cal B}} \!\D \vec{q}\:
\wit{C}(\vec{q},t) = \langle S_{\vec{x}}(t)^2\rangle = 1
\label{eq9}
\EEQ
This in turn fixes $\mathfrak{z}(t)$ or via (\ref{eq5}) the function
$g(t)$ as the solution of a nonlinear Volterra integral equation
\BEQ
g(t)= A(t) + 2T{\int_0^t\!\D t'\: f(t-t')g(t')} +
2 S_0 \int_{0}^{t}\!\D t'\, B_{\vec{x}}(t') \sqrt{g(t')}\: + 
\left( \int_{0}^{t}\!\D t'\, B_{\vec{x}}(t') \sqrt{g(t')} \right)^2
\label{eq15}
\EEQ
For $S_0=0$ and $T=0$, eqs.~(\ref{eq13},\ref{eq15}) had been derived  
before for the spherical spin-glass \cite{Cugl95b}. Besides on time, 
$g(t)$ also depends on the temperature
$T$ and the initial conditions parametrized by $S_0$ and $C_{\vec{x}}(0)$. 

The expressions for $A(t)$ and $B_{\vec{x}}(t)$ simplify in certain cases. 
For uncorrelated initial conditions
\BEQ
C_{\vec{x},\vec{y}}(0) = \left( 1 - S_0^2 \right)  
\delta_{\vec{x},\vec{y}} + S_0^2
\label{eq10}
\EEQ
Then $\wit{C}(\vec{q},0)=1-S_0^2 +(2\pi)^d \delta^d(\vec{q}) S_0^2$ and
\BEQ \label{gl:A}
A(t) = \left( 1 - S_0^2 \right) f(t) + S_0^2
\EEQ
For a spatially uniform magnetic field $H_{\vec{x}}(t) = H(t)$ we have
\BEQ \label{gl:B}
B_{\vec{x}}(t) = H(t)
\EEQ
These two conditions and consequently eqs.~(\ref{gl:A},\ref{gl:B}) will be used 
throughout this paper.

When $S_0\ne 0$, it will be useful to consider besides $C(t,s)$ also
the connected two-time autocorrelator (see \cite{Pico02} for an analogous
situation in the $1D$ Glauber-Ising model)
\BEQ \label{2:Gts}
\Gamma(t,s) = \langle S_{\vec{x}}(t) S_{\vec{x}}(s)\rangle - 
\langle S_{\vec{x}}(t)\rangle\langle S_{\vec{x}}(s)\rangle
\EEQ 
Finally, the response function is obtained in the usual way
\cite{Newm90,Kiss93,Cugl95a,Cugl95b,Godr00b,Pico02}
by considering the linear response to the magnetic field. 
It is easy to see that in Fourier space
\BEQ \label{2:Rqts}
\wit{R}(\vec{q},t,s) = \left.
\frac{\delta\langle\wit{S}(\vec{q},t)\rangle}{\delta \wit{h}(\vec{q},s)}
\right|_{h_{\vec{r}}=0}
= e^{-\omega(\vec{q})(t-s)} \sqrt{\frac{g(s)}{g(t)}}
\EEQ
{}From these expressions, the autocorrelation function
$C(t,s) = C_{\vec{0}}(t,s)$ and the autoresponse function
$R(t,s)=R_{\vec{0}}(t,s)$
can be obtained by integrating over the momentum $\vec{q}$.

Summarizing, the physically interesting correlation and response functions
are given by equations (\ref{eq13},\ref{2:Gts},\ref{2:Rqts}) together with
the constraint eq.~(\ref{eq15}). 
This constitutes the main result of the general formalism. 

In the next section, we turn towards the solution of these equations. Compared
to the case without an external magnetic field, this task is difficult since
the underlying Volterra equation (\ref{eq15}) is nonlinear. The mathematical
theory of nonlinear Volterra equations is still being developed \cite{Kara00}.
In a few cases, explicit analytic solutions can be found. Otherwise, we
shall turn to numerical methods.

\section{Solution of the constraint}

It is the peculiar feature of the kinetic spherical model that a complicated
many-body problem can be exactly reduced to the solution of a single
equation. We first derive the exact late-time asymptotic behaviour of 
the solution $g(t)$ of eq.~(\ref{eq15}) for a constant field $H(t)=H$,  
that is for times $t\gg 1$.
Afterwards, we comment on the use of asymptotic expansions for the 
calculation of physical observables.  

It is convenient to consider first the initial condition $S_0=0$ which is
easier to handle. As we shall see, the system actually looses its memory of 
the initial state quite rapidly. 

A first condition on the late-time asymptotics comes from the known fact
$\left|C_{\vec{x},\vec{x}}\right| \le 1$. Together with eq.~(\ref{eq13}),
it is easy to see that the power-law dependence of $g(t)$ on $t$ as found
\cite{Godr00b} for the special case $H=0$ and $T<T_c$ is incompatible with 
that condition in the case at hand.  

We therefore try, for late times $t\gg 1$, an asymptotic exponential ansatz 
\BEQ
g(t)=a\,e^{t/\tau},
\label{3:eq1}
\EEQ
where $a$ and $\tau$ are constants to be determined. 
Indeed eq.~(\ref{eq13}) now shows that $\left|C_{\vec{x},\vec{x}}\right|$ 
is bounded if $\tau>0$. To see this, observe that because of the ansatz 
(\ref{3:eq1}) the main contribution to the terms in eq.~(\ref{eq13}) which 
depend on the magnetic field comes from the upper limit of integration. Consequently,
the quadratic term in $B_{\vec{x}}(t)$ dominates over the terms linear in 
$B_{\vec{x}}(t)$ and also over those terms which do not contain 
$B_{\vec{x}}(t)$ at all. For large times $t,s$ we have asymptotically
\BEQ
\lim_{t,s \to \infty} C_{\vec{x},\vec{x}}(t,s)=4H^2\tau^2,
\label{3:eq2}
\EEQ
where the limit is taken for a constant time difference 
$\sigma=t-s \ge 0$. Inserting eq.~(\ref{3:eq1})
into eq.~(\ref{eq15}) yields for $S_0=0$, along the same lines 
\BEQ
g(t)= A(t) + 2T{\int_0^t\!\D t'\: f(t-t')g(t')} + 
      4 H^2 \tau^2 g(t).
\label{3:eq3}
\EEQ
and with $f(t)=(e^{-4t}I_0(4t))^d$ from eq.~(\ref{2:fAB}). 
Using the Laplace transformation
\BEQ
\lap{f}(p)=\int_0^\infty \,\D t \,f(t)\,e^{-pt}
\label{3:eq4}
\EEQ
we find from (\ref{3:eq3})
\BEQ
\lap{g}(p)=\frac{\lap{A}(p)}{1-2T\lap{f}(p)-4H^2\tau^2}
\label{3:eq5}
\EEQ
This must be 
consistent with the Laplace-transformed ansatz of eq.~(\ref{3:eq1})
\BEQ
\lap{g}(p)=\frac{a}{p-\tau^{-1}}.
\label{3:eq6}
\EEQ
These two expressions can only be compatible if the denominator in 
eq.~(\ref{3:eq5}) vanishes at $p=\tau^{-1}$, i.e.
\BEQ
1-2T\lap{f}(\tau^{-1}) = 4H^2\tau^2
\label{3:eq7}
\EEQ
and this must be a simple intersection (from eq.~(\ref{gl:A}) we know that 
$A(t)>0$, therefore $\lap{A}(p)>0$ can be related to $a$). Eq.~(\ref{3:eq7}) 
is an implicit equation for $\tau$ and we now show that there is always an
unique solution, provided $H\ne 0$. 

First, we consider the case $\tau\to 0$. From the definition of $\lap{f}(p)$
and $f(t)>0$, we have $\lap{f}(p)>0$. Similarly, $f(t)\leq 1$ for $t\geq 0$ 
implies $\lap{f}(p)\le p^{-1}$. Therefore 
\BEQ
\lim_{\tau\to 0}\left(1-2T\lap{f}(\tau^{-1})\right)=1
\EEQ
Second, we consider the case $\tau\to\infty$. From the results of 
\cite{Godr00b} on $\lap{f}(p)$ one has
\BEQ\label{3:eq8}
\lim_{\tau\to\infty} \left(1-2T\lap{f}(\tau^{-1})\right)=1-\frac{T}{T_c}
\EEQ
It is well-known that the Laplace transformation $\lap{f}(p)$ 
of a positive function $f(t)$ decreases monotonously with $p$ \cite{Abra65}. 
Therefore, the left-hand-side of eq.~(\ref{3:eq7}) 
decreases monotonously from 1 
to $1-T/T_c$ as $\tau$ increases from 0 to $\infty$, while the right-hand-side 
$\sim \tau^2$ increases monotonously for $H\ne 0$. 
This establishes the existence of a simple intersection and therefore of a 
unique $\tau$ which describes the
late-time asymptotics of $g(t)$ for $H\ne 0$, see eq.~(\ref{3:eq1}). 
For $\left|H\right| \to 0$ and $T\leq T_c$ we find
$\tau \to \infty$, while for $H=0$ a 
solution for $\tau$ only exists if $T>T_c$. This reproduces the well-known
result that for $H=0$, $g(t)$ only has an exponential behaviour for 
$T>T_c$ \cite{Godr00b,Pico02}. The fact that in the case
$H\neq 0$ we find an exponential behaviour for 
all temperatures $T$ shows that the system 
relaxes to an equilibrium state after the finite time $\tau$ \cite{Cugl95b} 
and neither critical behaviour nor ageing is expected for late times. 
It is now clear that adding the extra terms coming in for $S_0\ne 0$ will
merely generate sub-leading corrections and the asymptotic
solution eq.~(\ref{3:eq1}) will not be affected. 

In conclusion, we have established: {\it the leading long-time
asymptotic behaviour of $g(t)$ is given by eq.~(\ref{3:eq1}) where $\tau$ is
the unique solution of eq.~(\ref{3:eq7}) and with 
$a=-\lap{A}(1/\tau)/(2T\lap{f}'(1/\tau))$, 
for any value $H\ne 0$ of the constant magnetic field 
and any given mean initial magnetization $S_0$.} 

For finite times, there is no analytical solution of eq.~(\ref{eq15}) 
available. Instead, as described in appendix~A, we determine $g(t)$ 
numerically.  
Although the two--time observables are the relevant 
quantities for the study of ageing phenomena (see section~4), it is still
useful to consider single--time observables
like the average magnetization $S(t)$ given by
\BEQ
S(t)=\langle S_{\vec{x}}(t) \rangle = 
   \frac{1}{\sqrt{g(t)}}\left[ S_0 + 
   \int_{0}^{t}\!\D t^\prime H(t^\prime) \sqrt{g(t^\prime)}\right]
\label{3:eq9}
\EEQ

\begin{figure}[tn]
\centerline{\epsfxsize=3.5in\epsfbox
{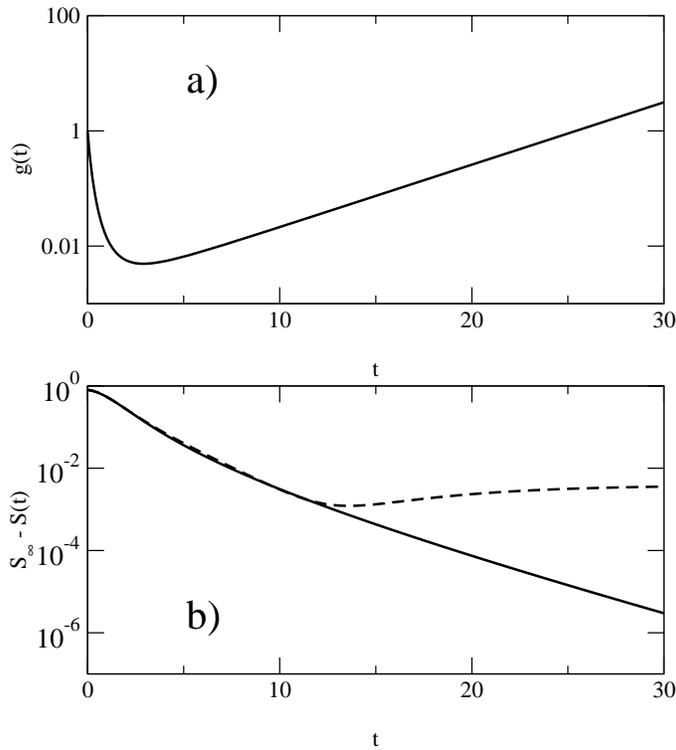}
}
\caption{The function $g(t)$ (a) and the distance of the magnetization to its
equilibrium value (b) for $d=3.5$, $T=2\quad 
(T_c\approx 5.27)$, $H=0.1$ and $S_0=0$.
The full curve shows the results for the direct numerical calculation, 
the dashed line shows the results for the 5th order fit which 
coincides with the full curve in (a). 
\label{Bild1}}
\end{figure}

In practise, care is required in using asymptotic solutions of
$g(t)$ for the prediction of the time-dependence of observables
such as $S(t)$. We illustrate this in figure~\ref{Bild1}, where
$g(t)$ and the distance of the magnetization to its
equilibrium value $S_\infty - S(t)$ are shown as a function of time. 
We see in figure~\ref{Bild1}a that after an initial fall-off, $g(t)$
quickly reaches the asymptotical regime of exponential growth. In
figure~\ref{Bild1}b, however, we compare the mean magnetization $S(t)$
as found from the exact numerical solution $g(t)$ (see appendix~A) with
the one obtained from an asymptotic fit of the form
\BEQ
g(t) \simeq e^{t/\tau} \sum_{\ell=0}^{\ell_{\rm max}} a_{\ell} (t-t_0)^{-\ell}
\EEQ
where we use $\ell_{\rm max}=5$. Although that asymptotic fit for $g(t)$
cannot be distinguished from the exact numerical result in figure~\ref{Bild1}a,
the deviation in $S(t)$ is considerable. 

In the rest of this paper, we shall use the direct numerical solution
of eq.~(\ref{eq15}). 

\section{Single-time observables}

Our first applications consider single-time observables, which are the 
ones most commonly studied. 
 
\begin{figure}[tn]
\centerline{\epsfxsize=3.5in\epsfbox
{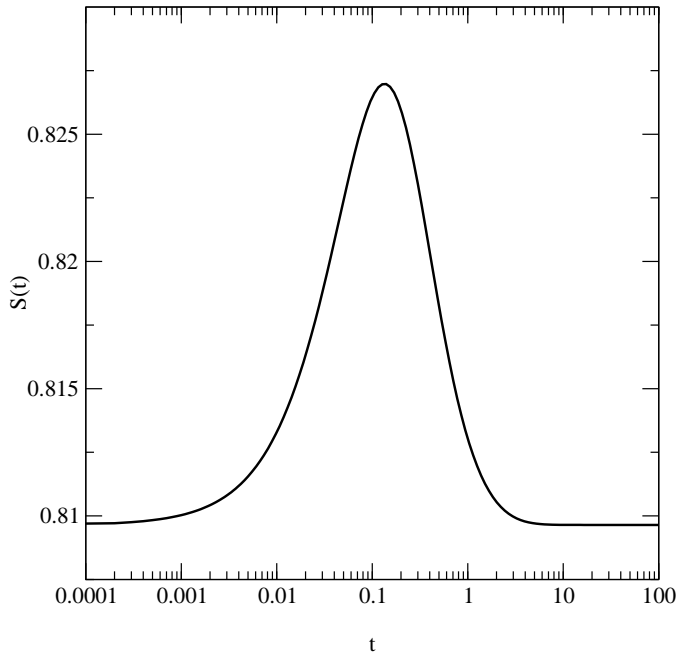}
}
\caption{The magnetization of the system evolving from 
$S_0=S_\infty$ calculated for 
$d=3.5$, $T=2(<T_c)$, $H=0.2$ and $S_0\approx 0.810$. 
\label{Bild2}}
\end{figure}

An instructive example on the importance of fluctuation effects
is constructed as follows. For a given external magnetic field $H$,
one may easily calculate the equilibrium magnetization $M_{\rm eq}$. 
Now prepare the system such that the spins have a mean magnetization
$M_{\rm eq}$ but such that spins on different sites are
uncorrelated. The time evolution of $S(t)$ is shown in figure~\ref{Bild2}. 
While a mean-field description would have predicted a constant
$S(t)$, we see that the magnetization is not constant but 
increases towards a peak before it falls back to the equilibrium
value $M_{\rm eq}$. Intuitively, we would expect that the individual spins
tend to align with the local magnetic field provided by their neighbours. 
Since initially $S_0=M_{\rm eq}>0$, one orientation is preferred with 
respect to the other one and domains oriented in parallel to $M_{\rm eq}$ will
grow preferentially. When the domains have grown 
large enough the influence of this effect
decreases and the system approaches quickly the 
equilibrium and the magnetization decreases 
again. This picture, although close in spirit to the Ising model with
its discrete spin variables, also works in the spherical model, in spite of
the fact that the interaction can be reduced to a free-field theory. The
remnant interaction between different spins provided by the spherical 
constraint is sufficient to achieve non-trivial correlations between different 
spins.

\begin{figure}[tn]
\centerline{\epsfxsize=3.5in\epsfbox
{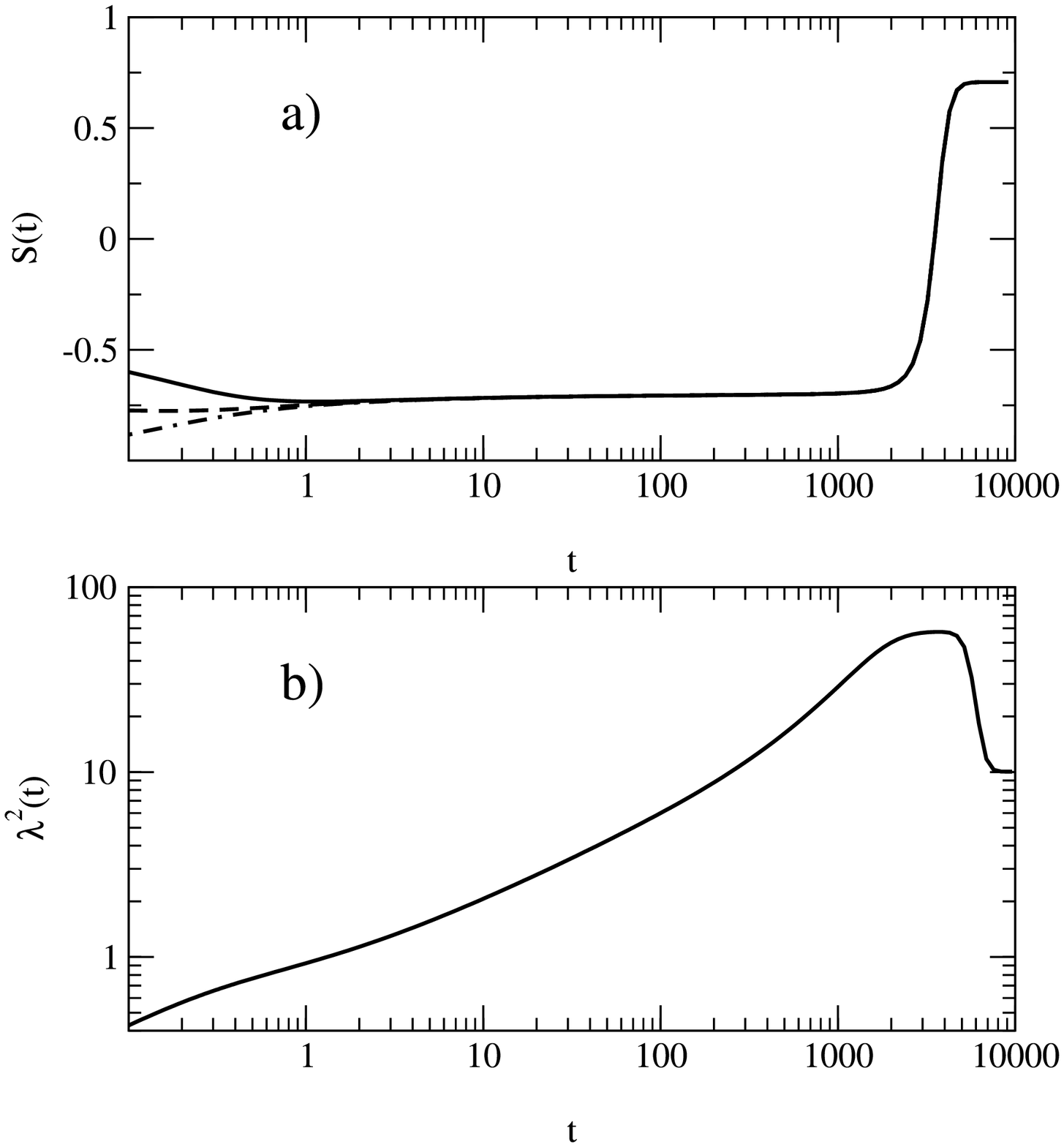}
}
\caption{(a) Average magnetization $S(t)$ for $d=3$, $T=2\: (<T_c\simeq 3.96)$,
$H=10^{-3}$ and $S_0=-0.5$ (full curve), $S_0=-0.75$ (dashed curve) and
$S_0=-1$ (dash-dotted curve). (b) Squared correlation  
length $\lambda(t)^2$ of 
fluctuations for $S_0=-0.5$, where the other parameters are as in (a).
\label{Bild3}}
\end{figure}

We have seen in the previous section that for $H\ne 0$ and very late times the
system relaxes back to its unique equilibrium state. For a vanishing magnetic
field, the {\rm equilibrium} free energy would have a double-well structure with
two equivalent minima, corresponding to the two possible orientations of the
mean magnetization. Turning on a magnetic field, this potential is tilted
and the depth of the two local minima is no longer the same. The lower
minimum becomes the unique equilibrium state, the other one corresponds to a
metastable state. It is clear that
if the system is initially prepared in the well corresponding to the 
equilibrium state, it will relax rapidly towards that state. Here we are  
interested how the transition from the metastable state towards the 
equilibrium state occurs. 

Therefore, we prepare the system with an initial magnetization antiparallel
to the given external field. In figure~\ref{Bild3}a we show the time evolution
of the mean magnetization. After a short time the system reaches the metastable 
state, independently of the absolute value of the initial magnetization $S_0$,
and where $S(t)$ stays practically constant. The system remains in the
metastable state for several decades until the magnetization
is reversed quite rapidly (although one should not be misled by the logarithmic
time-scale in this figure which makes the changeover 
to appear be very fast). In order to understand better what is going on
we define a characteristic length $\lambda(t)$ of the {\em fluctuations}
\BEQ
\lambda(t)^2=\sum_{\vec{r}\in\Lambda} \vec{r}^2 
\left(C_\vec{r}(t,t)-S(t)^2\right),
\EEQ
where $\vec{r}$ runs over all sites of the lattice $\Lambda\subset\mathbb{Z}^d$ 
and $C_\vec{r}(t,s)$ is the spin--spin correlation 
function (\ref{eq8}). The time evolution of $\lambda(t)$ is shown in
figure~\ref{Bild3}b. Starting from a very small initial value, $\lambda(t)$ 
increases towards a maximum value which is reached at the time when $S(t)$ 
starts to deviate perceptively from its value in the metastable state. 
While $S(t)$ changes its sign, $\lambda(t)$ remains approximately constant at
its maximal values before it relaxes towards the equilibrium correlation
length, with a typical value of a few lattice spacings. The coincidence
of the times of the reversal of $S(t)$ and the peak in $\lambda(t)$ shows
that whole domains rather than single uncorrelated spins are flipping.

\section{Two-time observables}

Having seen that the magnetization reversal passes via an intermediate
state with highly correlated fluctuations, we discuss in this section how this
manifests itself in the behaviour of the two-time quantities. An important
quantity is the time $\vartheta$ after which the magnetization reverses itself.
Evidently, $\vartheta=\vartheta(H,T,d)$, but we have not investigated in detail how
$\vartheta$ depends on these parameters in detail. For illustration
purposes, we shall use in this section the same choice of parameter values
as in figure~\ref{Bild3}, then $\vartheta\approx 3000$. 
For finite values of $t$, $g(t)$ can be readily found from eq.~(\ref{eq15}) 
using the numerical methods described in appendix~A. We shall focus on the 
metastable state by restricting to waiting times $s$ in the intermediate time 
regime $s\leq \vartheta$. A magnetization reversal is seen if the
initial magnetization is chosen antiparallel to the external field. 

Our choice of initially uncorrelated spin with a mean magnetization $S_0$
can be considered as a special case of initially correlated spins. The case
of spatial long-range correlations of the form 
$C_{\rm ini}(\vec{r}) \sim |\vec{r}|^{-d-\alpha}$ in the initial state 
was studied in detail before \cite{Newm90,Pico02}. Formally, this reduces
to an initial state with a constant mean magnetization in the limit 
$\alpha\to -d$. Using the exact results of \cite{Pico02} for $T<T_c$, we have 
\BEA
C(t,s)&=&1-\frac{T}{T_c}=M_{\rm eq}^2 \nonumber \\
R(t,s)&=&\left[4\pi(t-s)\right]^{-d/2},
\label{4:eq1}
\EEA  
where $M_{\rm eq}$ is the equilibrium magnetization.

\begin{figure}[tn]
\centerline{\epsfxsize=4in\epsfbox
{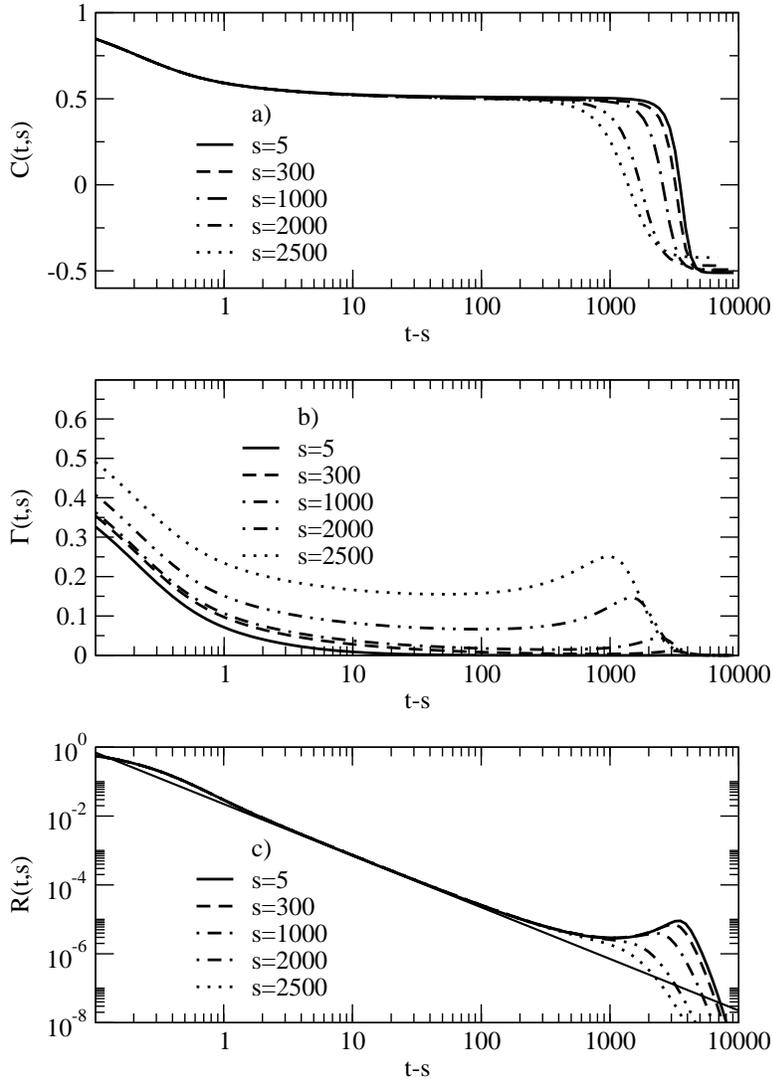}
}
\caption{The two--time autocorrelation function $C(t,s)$ (a), 
the correlation of fluctuations 
$\Gamma(t,s)$ (b) and the response function 
$R(t,s)$ (c) plotted vs. the time
difference $t-s$ for waiting
times $s=\{ 5;300;1000;2000;2500\} $; 
the data were calculated for $d=3$,
$T=2$ and $H=10^{-3}$. In (c) the straight 
line shows the formula $\left[ 4\pi(t-s)\right]^{-d/2}$.                        
\label{Bild4}}
\end{figure}

In figure~\ref{Bild4}a the correlation 
function $C(t,s)$ is plotted versus the time
difference $t-s$ for several values of the waiting times $s$ which are chosen 
to be in the metastable state, that is $s\leq \vartheta$ 
(compare figure~\ref{Bild3}a). After a short time
the curves reach a plateau, with a value very close to the equilibrium value 
$C(t,s)=M_{\rm eq}^2$ (a small contribution of the magnetic field can be 
neglected here). $C(t,s)$ maintains itself at this value for approximatively 
three decades, independently of the waiting time $s$. 
When the observation time $t$ becomes larger than the magnetization 
reversal time $\vartheta$, the correlation function $C(t,s)$ changes its sign 
because the spins at time $s$ before the reversal are anticorrelated to the 
spins at time $t$ after the reversal. However, we point out that the 
changeover takes more time when the waiting time $s$ is increased. 
The curves rapidly approach the expected value $-M_{\rm eq}^2$ because spins
of the metastable state are compared to the stable state. So we conclude that
the correlation function $C(t,s)$ is mainly determined by the value of 
magnetization.

While $C(t,s)$ measures the time-dependence of the autocorrelation of a
given spin, $\Gamma(t,s)$, see eq.~(\ref{2:Gts}), measures the fluctuations. 
This is shown in figure~\ref{Bild4}b. It can 
be seen that for waiting times $s\leq 1000$, $\Gamma(t,s)$ decreases fairly
rapidly as a function of the time difference $t-s$. In addition, a small
peak is observed in the region $t\approx\vartheta$. But for waiting times
closer to the magnetization reversal time $\vartheta$ (here for
$s=2000$ and $s=2500$), the fluctuations have become quite substantial
and show a larger peak around $(t-s)+s\approx \vartheta$. 
This may be viewed as another hint for the existence of correlated domains: 
as the spins inside of a domain are highly correlated a fluctuation of a spin 
within such a domain will cause other spins in the domain to follow this  
fluctuation. In turn, a side-effect of the enhanced correlations is
a longer lifetime of a spin fluctuation. 
After the magnetization reversal, $\Gamma(t,s)$ rapidly falls to zero.

Finally, in figure~\ref{Bild4}c the response function $R(t,s)$ is shown.  
First, we observe that for a time region
of at least two decades we recover eq.~(\ref{4:eq1}), which was derived
in \cite{Pico02} for the case without an external field. 
In this region translation invariance holds and hence no ageing occurs. 
The system behaves as if it were in equilibrium although it is only in a  
metastable state. Second, for observation times $t$ getting closer to the
reversal time $\vartheta$, the response function begins 
to deviate from this simple behaviour. We point out that the curves for 
all waiting times $s$ still collapse onto each other and that this deviation
occurs although $C(t,s)$ still has not appreciably changed away from
$M_{\rm eq}^2$. Third, for times $t\gtrsim \vartheta$ the dependence on the
waiting times becomes obvious before the response 
curves decrease very fast. This can be explained by considering
that the memory of perturbations is lost during 
the reversal from the metastable to the stable state. 

\begin{figure}[th]
\centerline{\epsfxsize=3.5in\epsfbox
{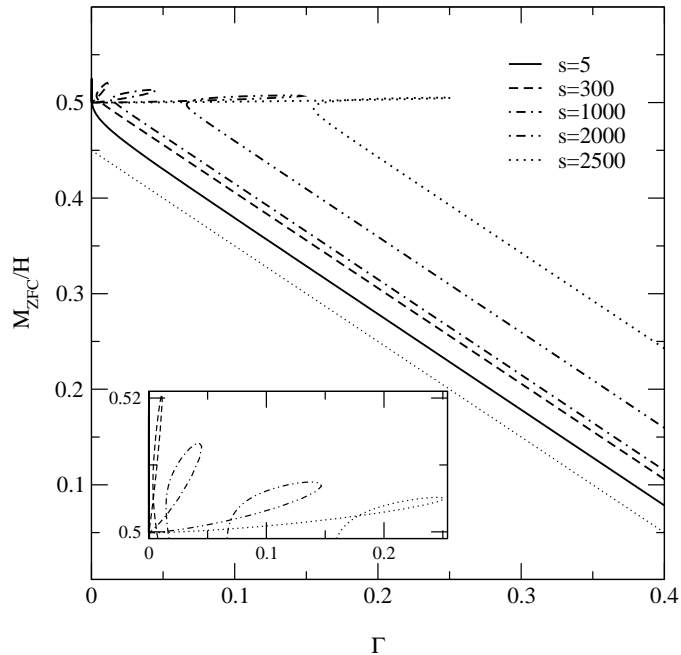}
}
\caption{$M_{\rm ZFC}/H$ vs. $\Gamma$, where the parameters are as in 
figure~\ref{Bild4}. For reference, the grey line gives the curve
$M_{\rm ZFC}/H=1\Gamma$. In the inset the region of the loops is
shown in more detail. 
\label{Bild5}}
\end{figure}

In order to decide whether the system is in 
equilibrium or not we shall investigate now
the zero--field--cooled (ZFC) magnetization which is defined by
\BEQ
M_{\rm ZFC}=H T \int_{s}^{t} \!\D u\, R(t,u).
\label{4:eq2}
\EEQ 
This quantity may be related to the fluctuation-dissipation 
ratio using eq.~(\ref{1:eq5}). Because of the non-vanishing initial
magnetization $S_0$ and the presence of an external magnetic field, the
quantities $C(t,s)$ and $\Gamma(t,s)$ are different and a 
fluctuation-dissipation
ratio is better defined using $\Gamma(t,s)$, namely
$X(t,s)=T R(t,s) \left( \partial\Gamma(t,s)/\partial s)\right)^{-1}$. This had
been checked explicitly in the $1D$ Glauber-Ising model \cite{Pico02} and in
certin simple model of glassy behaviour \cite{Buho02}. 
In spin glasses, it had been shown \cite{Cugl94a,Cugl94b} 
from mean-field theory that $X=X(C(t,s))$
although that is not necessarily so beyond mean field or in simple ferromagnets
\cite{Godr00b,Bert01,Chat03}. Nevertheless, this assumption is of good
heuristic value. In the spirit of the enterprise, let us consider the case
where here $X=X(\Gamma(t,s))$. This amounts to saying that $\Gamma$ serves as
a clock for the evolution of the system. Then 
\BEQ
M_{\rm ZFC}/H=\int_{\Gamma(t,s)}^{\Gamma(t,t)}\!\D \Gamma\, X(\Gamma).
\label{4:eq3}
\EEQ 
Consequently, when plotting $M_{\rm ZFC}(t,s)/H$ versus 
$\Gamma(t,s)$ for fixed $s$ (see figure~\ref{Bild5}) the slope of the curve 
corresponds to the value of $X$ -- provided of course that the assumptions
leading to (\ref{4:eq3}) are valid.\footnote{For metastable systems with
detailed balance and for time-scales shorter than the nucleation time, a  
fluctuation-dissipation relation is discussed in \cite{Baez03}.} 
Rather, we find in figure~\ref{Bild5} that with increasing waiting time $s$ 
the curves move from the lower right to the upper left. On the
other hand, for a given value of $s$, the system starts in the lower right
corner and moves rapidly along a curve $M_{\rm ZFC}(\Gamma)=\Gamma_0 -\Gamma$ 
until the metastable value $M_{\rm ZFC}/H=1-M_{\rm meta}\simeq 1-M_{\rm eq}$ 
is reached. The slope of unity of this curve is the same as would be found
for an equilibrium system. Surprisingly, while it undergoes the
magnetization reversal, the system then passes through a loop, which corresponds
to the peak in $\Gamma(t,s)$,
before it reaches a horizontal line, of height $1-M_{\rm eq}$. 
The movement along the horizontal line is a behaviour typical of the 
low-temperature phase, indeed through the
magnetization reversal the system behaves as if the quasiequilibrium branch
close to the metastable state had to be joined with the low-temperature
behaviour after the magnetization reversal. All in all, this behaviour is
quite analogous to the one observed for $M_{\rm ZFC}$ as a function of $C(t,s)$
for systems brought into the two-phase region by a temperature 
quench \cite{Pari99}.

Of course, all the results in this section depend on having taken 
$s\leq\vartheta$. If we take instead $s>\vartheta$, the system quickly relaxes
to its unique equilibrium state. 

\section{Dynamics in an oscillating field}

Further aspects of the magnetization reversal transitions become apparent
when the response of the system to a time-dependent external field $H=H(t)$
is studied. This allows to study hysteresis effects -- related to the easily
measured Barkhausen noise -- and has been studied for a long time, see 
\cite{Seth97,Rikv01} for reviews. From mean-field descriptions 
\cite{Tome90,Mend91,Fuji01}, one finds evidence that, depending on the 
amplitude and the period $P$ of $H(t)$, the time-dependent (and periodic) 
magnetization $S(t)=S(t+P)$ 
changes between two different forms.
First, there is a single symmetric solution (corresponding to the 
paramagnetic phase) such that
\BEQ \label{gl:6:1}
S(t+P/2) = - S(t). 
\EEQ
Second, there may exist a pair of non-symmetric solutions in the 
ferromagnetic phase where (\ref{gl:6:1}) does not hold. Indeed, the 
existence of a dynamical phase transition was established beyond mean-field
theory through simulations in the $2D$ Ising model with Glauber dynamics
\cite{Side99,Korn00,Korn02}. The order parameter of this transition
is the period--averaged magnetization $Q=Q(t)$ defined as
\BEQ
Q(t)=\frac{1}{P}\int_{t-P}^{t}\!\D t^\prime S(t^\prime),
\label{5:eq1}
\EEQ
where $P$ is the period of $H(t)$. In the Ising model for sufficiently strong 
fields and/or low frequencies $Q=0$ and $S(t)$ oscillates around zero,  
but $Q$ remains finite for smaller fields and higher frequencies and $S(t)$ 
then oscillates around one of the two values of the
equilibrium magnetization. Detailed finite-size scaling analysis has shown
that the exponents of $Q(t)$ and also of the associated susceptibility agree
with those of the {\em equilibrium} phase transition of the $2D$ Ising model
\cite{Side99,Korn00,Chat02}. This was further backed up by showing that the
equation of motion of the order parameter reduces to the $\phi^4$-theory
with noise \cite{Fuji01}
(similar studies were also performed on the equation of motion of the
anisotropic XY model \cite{Yasu02}).

Still, this kind of non-equilibrium phase transition need not generically 
exist. In the $q$-states Potts model with $q\geq 3$, for example, a mean-field
analysis shows that the 
time-dependent order parameter undergoes a cascade of period-doubling
bifurcations, rather than a simple phase transition \cite{Mend91}. It is
therefore of interest to explore the role of the topology of the phase
space further by considering a model in a different equilibrium universality
class.

\begin{figure}[t]
\centerline{\epsfxsize=3.5in\epsfbox
{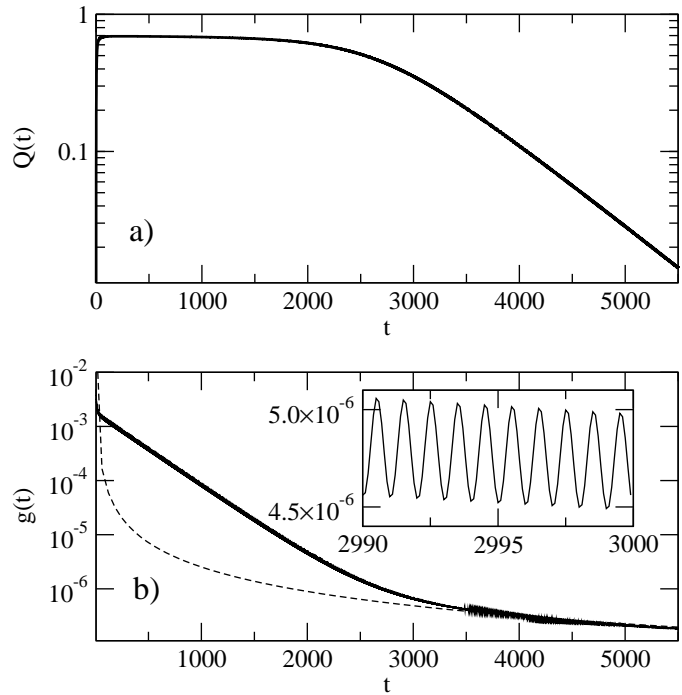}
}
\caption{(a) The period-averaged magnetization $Q(t)$ for 
$d=3$, $T=2$, $S_0=0$,
a sinusoidial external field with period $P=1.5$ and amplitude $H_0=0.2$.
(b) The Lagrange multiplier $g(t)$. On this scale only the behaviour of the 
bounds can be seen between which oscillations take place; these
oscillations are shown in the inset. The dashed line shows a power law 
$g(t)=0.1\, t^{-1.54}\approx c\cdot t^{-d/2}$. 
\label{Bild6}}
\end{figure}

\subsection{Behaviour of the magnetization}
We consider the spatially constant but time-dependent external field
\BEQ
H(t)=H_0 \sin\left(\frac{2\pi}{P} t\right).
\label{5:eq2}
\EEQ
The calculation of the observables follows the same lines as in the case of 
the constant field although a larger numerical effort is required. 
By inserting eq.~(\ref{3:eq9}) into eq.~(\ref{5:eq1}) the period-averaged 
magnetization $Q(t)$ is readily obtained. 
In figure~\ref{Bild6}a a typical example for $Q(t)$ is shown but the behaviour
seen in this case turns out to be generic. Taking the
Ising model as a guide, a heuristic argument \cite{Mend91} suggests that
a dynamic phase-transition should occur at least for all temperatures and
field amplitudes $H_0$ for which a metastable state exists. We therefore
used the same values for $T$ as before. 
However, in the spherical model we find that for small times $Q(t)$ takes
a plateau value before it decays exponentially for later times. In principle,
and in analogy with the Ising model, 
one might try to find the dynamic phase transition by measuring the time  
$\tau=\tau(P,H_0)$ when the transition between the plateau and the decay 
occurs. Following the practical experience of the Ising model either the scale 
of $H$ or $P$ can be normalized away, see \cite{Side99,Korn00,Korn02}. 
It should therefore be enough to vary the period $P$ (or the frequency) 
but keep the amplitude $H_0$ constant and map out $\tau(P,H_0)$. If there
is a dynamic phase transition at some critical period $P_c$, the cross-over
time should diverge $\tau(P_c,H_0)=\infty$. In practice, however, this
method is quite slow, because the calculations have to be done for 
increasingly larger time-scales.

It is a lot more efficient to study the Lagrange multiplier 
$g(t)$ which is shown in figure~\ref{Bild6}b. We observe that the value of
$g(t)$ oscillates between two bounds and the temporal behaviour of the
bounds correlates with the time-dependence of $Q(t)$. Namely, when
$Q(t)$ displays a plateau, the bounds for $g(t)$ decay exponentially with time 
while in the region of the exponential decay of $Q(t)$ the bounds of $g(t)$ 
decay with according to a power law. Therefore, the cross-over time 
$\tau(P,H_0)$ can be found by determining
the intersection of the two regimes for the bounds for $g(t)$.

\begin{figure}[t]
\centerline{\epsfxsize=3.5in\epsfbox
{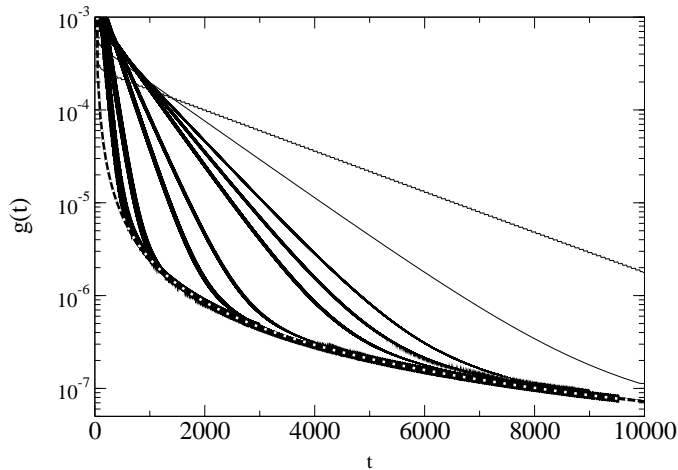}
}
\caption{$g(t)$ for $d=3$, $T=2$ and a sinusoidial external field with 
$H_0=0.2$; the periods are 
$P=\left\{0.4; 0.6; 0.8; 0.9; 1; 1.5; 2; 5; 10\right\}$ 
(from top to bottom). The dashed line shows a function 
$g_{\rm mas}(t) = 0.1\, t^{-1.54} \approx c\cdot t^{-d/2}$
\label{Bild7}}
\end{figure}

In figure \ref{Bild7} we display a typical behaviour of $g(t)$ for several 
values of the period $P$. We observe the cross-over from a roughly exponential 
behaviour $g_{\rm exp}(t)\approx \exp(-t/\mathfrak{t})$ with a relaxation 
time $\mathfrak{t}$ towards a master curve
$g_{\rm mas}(t)\sim {t}^{-1.54}$ which is reached for all given 
values of $P$ for sufficiently long times. In principle, one might try to 
estimate the time of cross-over between these two regimes by looking for the 
intersection of $g_{\rm exp}(t)$ and $g_{\rm mas}(t)$ and then further ask 
when this cross-over time will diverge in order to find the critical period 
$P_c$. Since for finite $\mathfrak{t}$ this intersection will always occur, 
a more reliable estimate of $P_c$ will be given by the condition 
$\mathfrak{t}^{-1}(P_c,H_0)=0$.

\begin{figure}[th]
\vspace{2mm}
\centerline{\epsfxsize=3.5in\epsfbox
{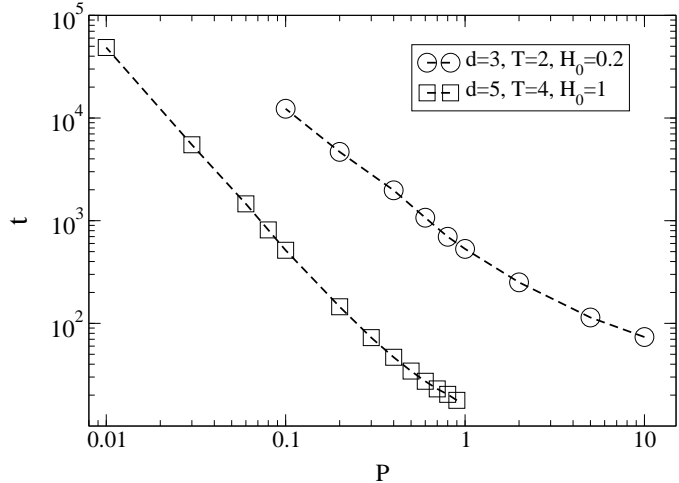}
}
\caption{$\mathfrak{t}(P,H_0)$ for $d=3$ and $d=5$ as read off from 
$g_{\rm exp}(t)$, see text. 
\label{Bild8}}
\end{figure}

In figure \ref{Bild8} we show $\mathfrak{t}(P,H_0)$ for $d=3$ and $d=5$, 
that is below and above the upper
critical dimension of the equilibrium critical behaviour. 
In all curves, we see that
$\mathfrak{t}(P)$ remains finite for all values of $P$ which we considered.
Phenomenologically, $\mathfrak{t}\sim 1/P^{v}$ for $P$ small enough and 
some exponent $v>0$ ($v\approx 1.4$ in $3D$; $v\simeq 1.95$ in $5D$). 
The fact that $\mathfrak{t}$ only diverges as $P\to 0$ is evidence that there 
is {\em no DPT} in the spherical model in an oscillating magnetic field,  
in contrast to established results \cite{Side99,Korn00,Korn02,Chat02}
in the 2D Ising model and also with results on the $n\to\infty$ limit
of the O($n$) model \cite{Rao90,Dhar92}. 
We also see from figure~\ref{Bild8} that the absence of 
the DPT is not related to whether or not the equilibrium phase transition of 
the spherical model is in the mean-field regime.

\begin{figure}[t]
\centerline{\epsfxsize=3.5in\epsfbox
{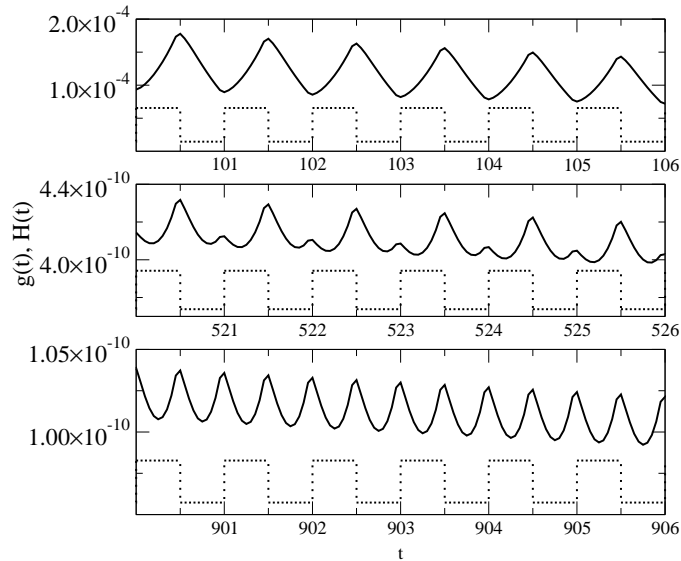}
}
\caption{$g(t)$ (full line) compared to $H(t)$ 
(dotted line, scaled and shifted) 
for different times. These calculations were done 
for $d=5$, $T=4$, $S_0=0$ and a rectangular external field with amplitude 
$H_0=0.6$ and with period $P=1$.
\label{Bild9}}
\end{figure}

\subsection{Behaviour of the Lagrange multiplier}

By fitting $g_{\rm mas}(t)$ for $d=3$ and $d=5$ we find exponents of  
$w=1.52\pm 0.01$ and $w=2.51\pm 0.01$ respectively. From these observations,
we conjecture that for sufficiently long times, the Lagrange multiplier
$g(t)$ satisfies the bounds
\BEQ \label{6:Schranken}
C_1 \leq t^{w} g(t) \leq C_2
\EEQ
with an exponent $w=d/2$ and some constants $C_{1,2}$. 
Indeed, we have also checked that 
these bounds hold not only for sinusoidal fields $H(t)$, but for
triangular and rectangular oscillating fields as well. 
Remarkably, the conjectured exponent $w=d/2$ of the power-law bounds 
eq.~(\ref{6:Schranken}) coincides with the same value found for the kinetic
spherical model {\em without} a magnetic field \cite{Godr00b}~! 
We have checked this for several values of the dimension $d$ and 
temperatures $T>0$. In appendix~B,
we derive the bounds (\ref{6:Schranken}) and especially the exponent $w=d/2$ 
in the $P\to 0$ limit and for $T=0$, under mild additional conditions.  
A fully disordered initial state simplifies
the calculations but the result remains the same for any short-ranged initial
correlators. Therefore, the relaxation time $\mathfrak{t}(P,H_0)$ is formally
infinite for $P\ll 1$ and $T=0$. We have thus shown the {\em absence
of a DPT in the physical situation where it would have been expected to be seen
first}. In this respect, the spherical model behaves in quite a different
way than the Ising model. The rigourous derivation
of eq.~(\ref{6:Schranken}) is left as an open mathematical problem.

The absence of a dynamical phase transition is further illustrated in
figure~\ref{Bild9}. There we compare $g(t)$ with a rectangular field $H(t)$ 
(scaled and shifted for convenience). While for small times, $g(t)$ oscillates
with the driving period $P$, we see that with $t$ increasing, an additional
peak builds up until $g(t)$ oscillates with half the period of the driving
field at late times. The fact that $g(t)$ oscillates with half the external
period $P$ is an indication that the system is described by the symmetric
solution, see eq.~(\ref{gl:6:1}). 
The same kind of period-halving is also found for triangular and sinusoidal
fields. 

\begin{figure}[t]
\centerline{\epsfxsize=3.5in\epsfbox
{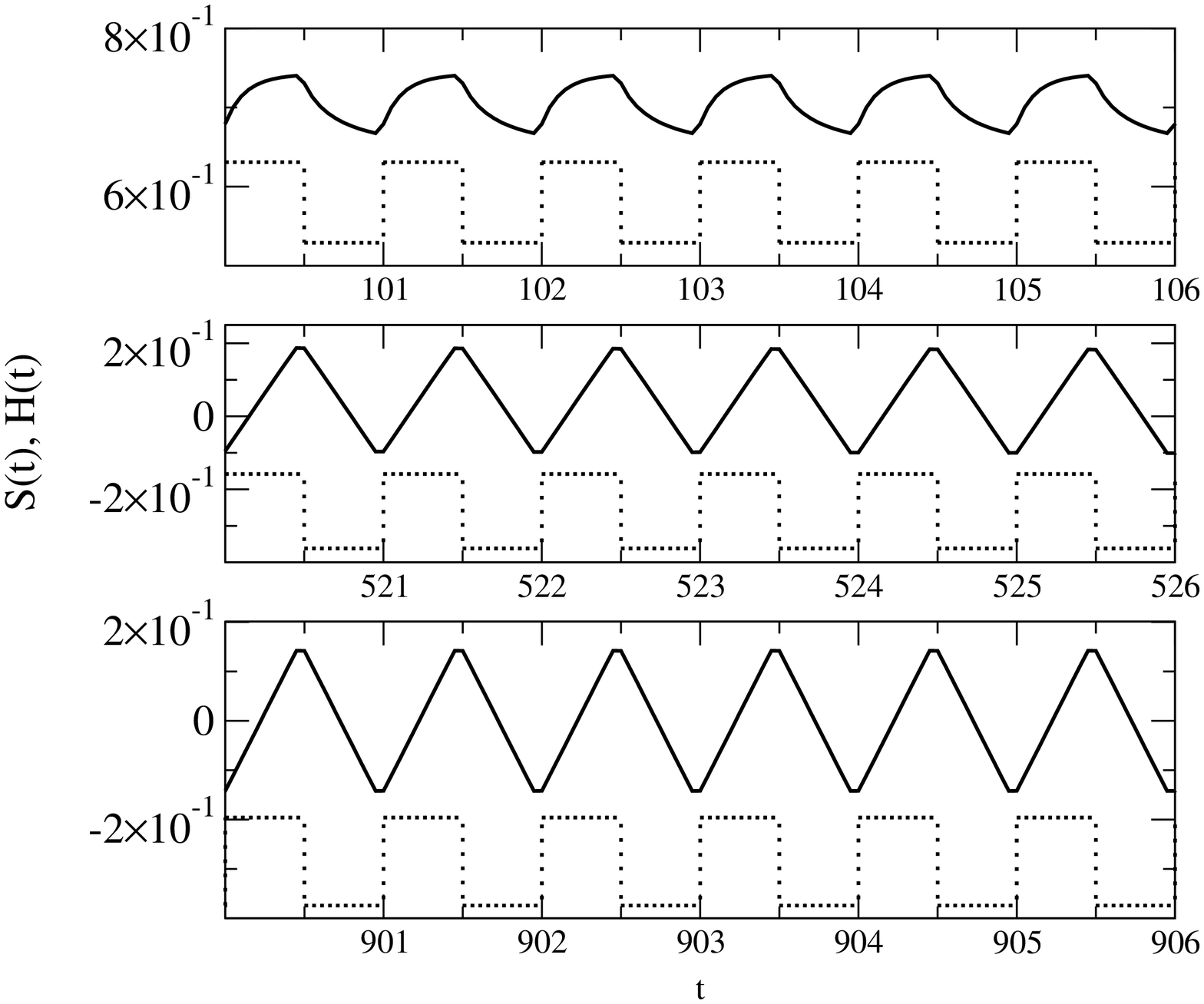}
}
\caption{Time evolution of the mean magnetization $S(t)$ (full line) compared 
to the one of a rectangular magnetic field $H(t)$ 
(dotted line, scaled and shifted) 
for different time regimes; the parameters are as in fig.~\ref{Bild9}. 
\label{Bild10}}
\end{figure}

This phenomenon is easily understood:
since for small times the magnetization oscillates around a non-vanishing value,
the global symmetry is broken and the two half-periods of the external field 
affect the system in two qualitatively different ways. However, for later times 
the magnetization oscillates around zero and there is no qualitative difference 
of the response of the system between the
two half-periods of the external field any more. This fact is reflected by 
$g(t)$ actually becoming periodic with period $P/2$, viz. $g(t+P/2)=g(t)$. 

\begin{figure}[th]
\centerline{\epsfxsize=3.5in\epsfbox
{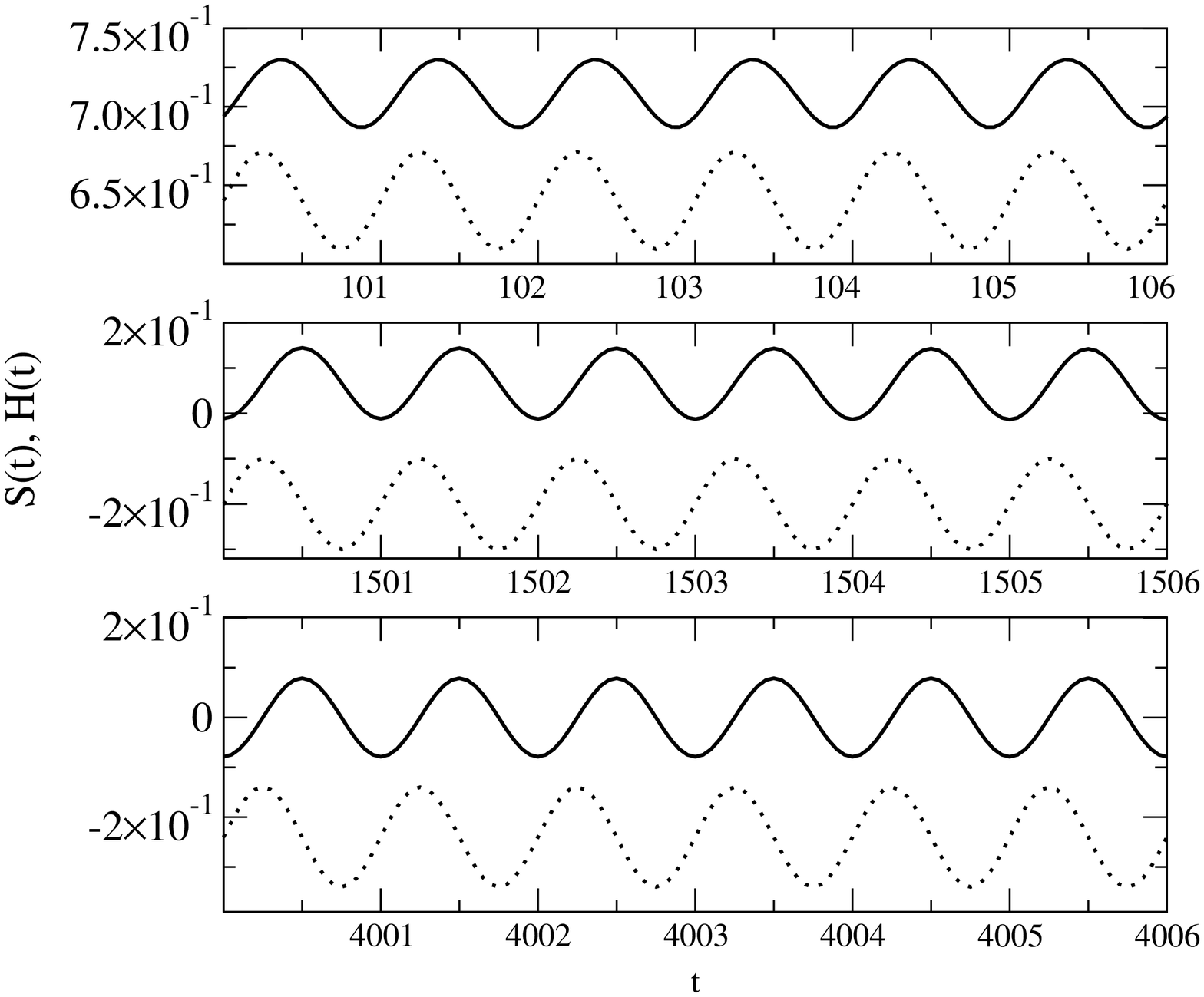}
}
\caption{Time evolution of the mean magnetization $S(t)$ (full line) compared 
to the one of a sinusoidal magnetic field $H(t)$ with amplitude $H_0=0.5$
(dotted line, scaled and shifted) 
for different time regimes; the other parameters are as in fig.~\ref{Bild9}. 
\label{Bild11}}
\end{figure}

\begin{figure}[th]
\centerline{\epsfxsize=3.5in\epsfbox
{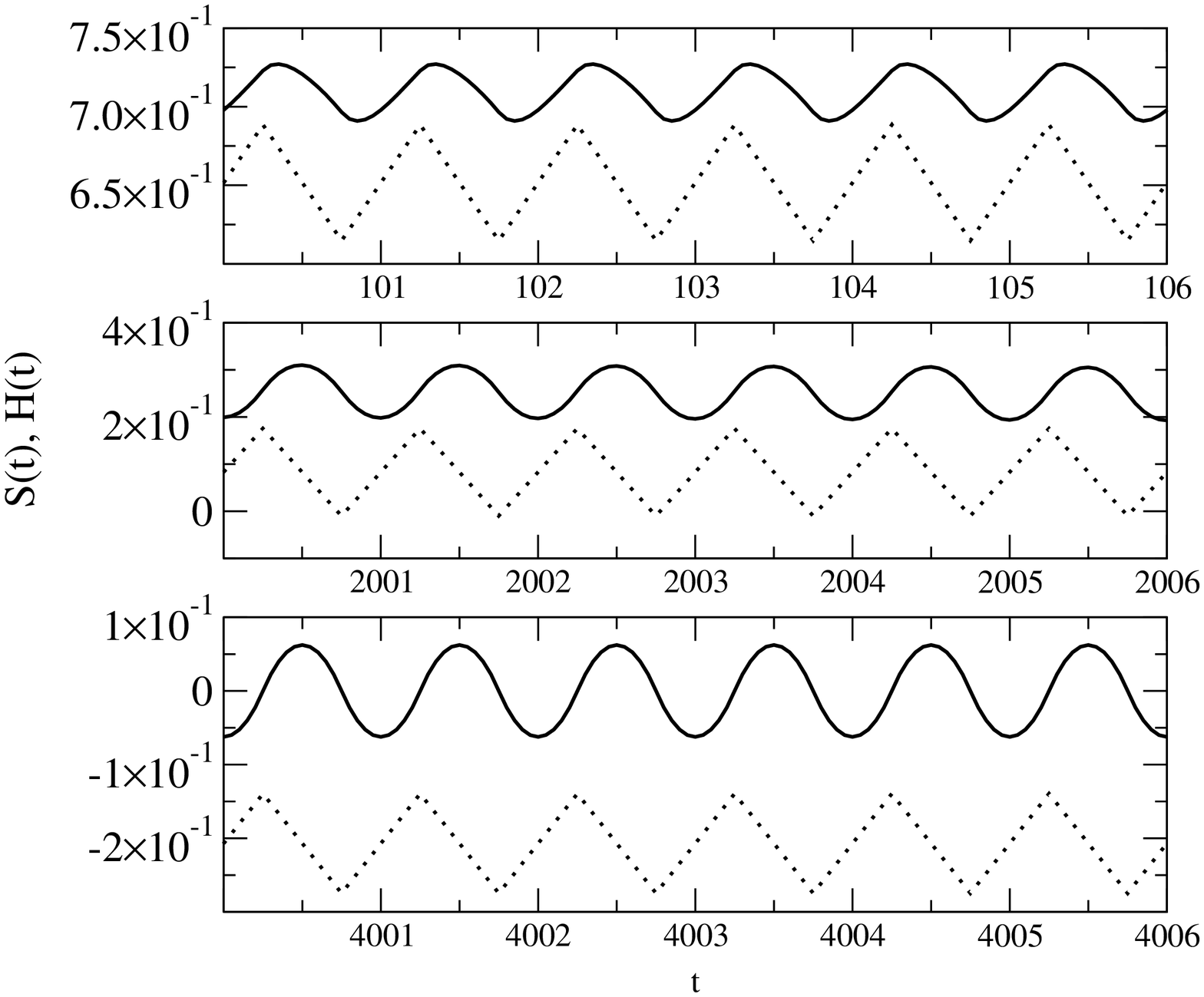}
}
\caption{Time evolution of the mean magnetization $S(t)$ (full line) compared 
to the one of a triangular magnetic field $H(t)$ with amplitude $H_0=0.5$
(dotted line, scaled and shifted) 
for different time regimes; the other parameters are as in fig.~\ref{Bild9}. 
\label{Bild12}}
\end{figure}
 
The behaviour of $S(t)$ is illustrated in figures~\ref{Bild10}, \ref{Bild11}, 
and \ref{Bild12}. For relatively
small times (upper panel), $S(t)$ oscillates around the positive equilibrium 
value and is periodic with period $P$. It is interesting to note that the
qualitative shape of $S(t)$ for the rectangular oscillating external field 
in this regime matches closely the one observed
in the dynamically ordered phase of the $2D$ Ising model, 
see \cite[figure 2(b)]{Korn00}. For larger times, the dynamic order
parameter $Q(t)$ decreases until $S(t)$ oscillates around zero. The slow
cross-over towards a solution which satisfies eq.~(\ref{gl:6:1}) is illustrated
in the middle panels of figures~\ref{Bild10}-\ref{Bild12} and in the lowest
panels, a situation near to (\ref{gl:6:1}) is reached, where $S(t)$ becomes
{\em anti}periodic with period $P/2$. In the case of a rectangular field shown 
in figure~\ref{Bild10} the external 
magnetic-field amplitude is still rather small which results in a 
linear increase and decrease of the magnetization. For stronger fields the 
magnetization reaches saturation during one half--period and the behaviour 
of $S(t)$ deviates from piecewise linearity. The comparison to the sinusoidal
and triangular oscillating external field shows that in all three cases
the magnetization follows the integrated external field for not too 
large amplitudes.

The main result of this section is surprising: in spite of the fact that for 
$T<T_c$ there are just two equilibrium states for both the Ising and the  
spherical models in a (sufficiently small) constant magnetic field, the 
well-established dynamic phase-transition of the Ising model in a temporally 
oscillating magnetic field is apparently {\em absent in the spherical model}. 

\section{Conclusions}

In this paper we have investigated the non-equilibrium behaviour of the
spherical model in an external magnetic field. 
The model's dynamics is described in terms of a Langevin
equation and all quantities of physical interest can be expressed exactly in 
terms of the solution of a non-linear Volterra integral equation. 
In few especially simple cases that Volterra equation can be solved exactly, 
but we have in general used numerical methods. 

First, we studied the magnetization reversal transition, in a temporally  
constant magnetic field, which occurs if the
system is initially prepared in near to metastable state from which it
relaxes towards to unique equilibrium state. We find that the system 
evolves into the metastable state quickly and remains there for considerably 
long times until it finally relaxes into the stable state. For not too
large magnetic fields, this transition passes through transient states 
with long-ranged correlations of fluctuations, which means that during the
magnetization-reversal transition 
whole domains rather than single uncorrelated spins turn over.   

The two-time autocorrelation function is mainly determined by the 
magnetization so that connected correlation functions, which are more sensible 
to fluctuations, reveal more information. Again we find
that the transition involves long-ranged correlations. 
For times smaller than the transition time $\vartheta$ we find an effective 
equilibrium behaviour although the system is merely in the metastable state. 
In many respects, notably the fluctuation-dissipation relations, 
we find a close analogy with the ageing behaviour encountered 
in the absence of an external field. But approaching the 
magnetization-reversal the autoresponse function and the 
fluctuation-dissipation ratio show unusual behaviour, indicating that the 
process is rather complex. Therefore, although the non-vanishing magnetic
field $H$ sets a finite time scale for the relaxation towards the single
equilibrium state, we have found a very rich transient behaviour which in many
respects is quite analogous to the true ageing behaviour found without an
external field. 

Second, we looked for a dynamic phase transition in a time-dependent external 
magnetic field $H(t)$. Surprisingly, we find evidence that 
a dynamic non-equilibrium phase transition, which is known to occur e.g. in the 
Ising model, apparently does not exist in the spherical model. 
For sufficiently low temperatures, we rather find that although the dynamic 
order parameter $Q(t)$ reaches a plateau value for small times, there is always 
a cross-over to a late-time regime where $Q(t)$ decays away to zero. On a
technical level, this finding can be represented through the conjecture
eq.~(\ref{6:Schranken}) which points to an unexpected similarity with the 
phase-ordering kinetics of the {\em zero-field} spherical model. 

Given that several equilibrium properties of the isotropic O($3$) 
Heisenberg model are closer to the ones of the spherical model than they
are to the Ising model (see introduction), our results raises the question
whether a dynamic phase transition for the {\em isotropic} O($3$)
Heisenberg model in an oscillating field exists.\footnote{Existing articles
on the DPT in Heisenberg models are either mean-field studies \cite{Turk03} or
consider the anisotropic case \cite{Jang01} (which should be more Ising-like).
One might anticipate the existence of a critical $n_c$ such that in the
O($n$)-model in an oscillating field, there is a DPT for $n<n_c$ analogously 
to the Ising model and none for $n>n_c$.}

Lastly, our results beg the questions what are the effects of a magnetic field
on the kinetics of a spin-glass and what becomes of the magnetization reversal 
transition and the dynamical phase transition ? 
However, because of the well-known equivalence \cite{Zipp00} between the  
spherical spin-glass and the spherical ferromagnet, studies in different  
systems with a true glassy behaviour\footnote{The equilibrium behaviour of the
Ising spin glass in a magnetic field has been studied in detail, see
\cite{Pime02} and references therein. For the spherical spin glass in an
oscillating magnetic field short-time numerical calculations give
evidence in favour of a dynamic phase transition {\em at} 
$T=0$ \cite{Bert01a}.} 
are needed to shed light on this issue. 

\newpage 
\noindent {\large\bf Acknowledgements}\\

\noindent
We thank P Rikvold, JP Garrahan and L Berthier for useful correspondence. 
MP is grateful to the Deutscher Akademischer Austauschdienst (DAAD)
for financial support (DAAD Doktorandenstipendium im Rahmen des gemeinsamen 
Hochschulsonderprogramms III von Bund und L\"andern).

\appsection{A}{Numerical method}

We briefly discuss the numerical solution of the nonlinear Volterra equation
(\ref{eq15}), adapting standard methods \cite{Pres92} to the case at hand. 

Eq.~(\ref{eq15}) is cast into the following form, using the equations 
(\ref{gl:A}) and (\ref{gl:B}) 
\BEQ
g(t)= (1-S_0^2)f(t)+S_0^2 + 2T{\int_0^t\!\D t'\: f(t-t')g(t')} +
2 S_0 \int_{0}^{t}\!\D t'\, H(t') \sqrt{g(t')}\: + 
\left( \int_{0}^{t}\!\D t'\, H(t') \sqrt{g(t')} \right)^2.
\label{APP:eq1}
\EEQ
As a first step we will discretize the time by dividing the time interval
in $N-1$ segments of length $k$
\BEQ
t_i=k\,i \;\; ; \;\; i=0,1,\ldots,N-1.
\label{APP:eq2}
\EEQ
The continuous functions $f(t)$ are replaced by the $N$ dimensional vectors 
\BEQ
\vec{f}=(f_0,f_1,\ldots,f_N-1)^T \;\; , \;\; f_i=f(t_i)
\label{APP:eq3}
\EEQ
and the integrals are replaced by a sum by means of the extended
trapezoidal rule \cite{Pres92}
\BEQ
\int_{x_0}^{x_{N-1}} \!\D x f(x) \approx k
\left[ \frac{1}{2}f_0+f_1+f_2+\ldots+f_{N-2}+ \frac{1}{2}f_{N-1}\right].
\label{APP:eq4}
\EEQ
Therefore, we have the set of equations
\BEA
F_0\left(\left(g_0,\sqrt{g_0}\right),\vec{f},\vec{H},k\right)&=&0 \nonumber \\
F_1\left( \left(g_0,\sqrt{g_0}\right), 
\left(g_1,\sqrt{g_1}\right),\vec{f},\vec{H},k\right) &=&0 \nonumber \\
&\ldots& \nonumber \\
F_{N-1}\left( \left(g_0,\sqrt{g_0}\right), 
\left(g_1,\sqrt{g_1}\right), \ldots, 
\left(g_{N-1},\sqrt{g_{N-1}}\right),\vec{f},\vec{H},k\right) &=&0,
\label{APP:eq5}
\EEA
depending on the known vectors $\vec{f}$ and $\vec{H}$ and the step size $k$. 
We have $F_0=g_0-1$ and 
\BEA
\lefteqn{ 
F_i\left(\left(g_0,\sqrt{g_0}\right), \ldots, 
\left(g_{i},\sqrt{g_{i}}\right),\vec{f},\vec{H},k\right)= g_i 
\left[ Tkf_0+\frac{1}{4}k^2 H_i^2-1 \right] } 
\nonumber \\
&+& \sqrt{g_i}  \left[ S_0kH_i+k^2\left(\frac{1}{2}H_0\sqrt{g_0}+
\sum_{j=1}^{i-1} H_j\sqrt{g_j}\right) H_i \right] +
\left(1-S_0^2\right)f_i+S_0^2+2Tk  
\left( \frac{1}{2} f_i g_0 + \sum_{j=1}^{i-1}f_{i-j}g_j\right) 
\nonumber \\
& +& 2S_0 k \left( \frac{1}{2} H_0 \sqrt{g_0} +  
\sum_{j=1}^{i-1} H_j\sqrt{g_j}\right)+ 
k^2 \left(\frac{1}{2}H_0\sqrt{g_0}+
\sum_{j=1}^{i-1}H_j\sqrt{g_j}\right)^2
\label{APP:eq6}
\EEA
This set of equations
can be solved iteratively: $F_0$ determines 
$g_0$, $F_1$ then leads to $g_1$ and so on.  
However, since the $F_i$ are functions of $g_i$ and 
$\sqrt{g_i}$ at each step of iteration
two {\it a priori} distinct solutions for the $g_i$ are found. They may be
obtained by replacing 
$G_i=\sqrt{g_i}$ and solving the resulting
quadratic equation in $G_i$. So the question arises 
which of these solutions has to be used.

We calculated the two solutions for $\vec{g}$ when 
only using either the solution
according to the positive root (`+'--curve) or the 
negative one (`--'--curve) ---
the exact solution should evolve somewhere between 
these two limiting curves. 
We found that decreasing the step size $k$ results 
in an approach of the `--'--curve 
to the `+'--curve where the latter one only slightly 
changes. Finally choosing a 
sufficient small $k$ the two curves collapse, 
so that the exact solution is found.
For larger values of $k$ the `+'--curve shows 
only small deviations to the limiting
curve, so that in all calculations this solution was used.

For all calculations a step size of $k=10^{-2}$ 
was sufficient, except for the data shown
in Fig.~\ref{Bild2} where $k=10^{-4}$ was used 
because there the time scale is much smaller.

The evaluation of the one-- or two--time observables 
proceeds by a straightforward
implementation of their defining integrals by the extended trapezoidal rule.

\appsection{B}{}

We derive the bounds eq.~(\ref{6:Schranken}) for the Lagrange multiplier  
$g(t)$ in an oscillating magnetic field $H(t)$, 
for the special case of vanishing temperature $T=0$. For convenience, 
we consider a fully disordered initial state and vanishing
initial magnetization $S_0=0$, but our results also hold true for
arbitrarily short-ranged initial conditions and $S_0\neq 0$. 
We define $G(t) := \sqrt{g(t)\,}$. 
The non-linear Volterra integral equation then is
\BEQ \label{gl:B:1}
G(t)^2 = A(t) + \left( \int_{0}^{t} \!\D s\, H(s) G(s) \right)^2
\EEQ
We shall assume throughout that $G(t)$ is bounded on the positive real
axis, that is $|G(t)|\leq M <\infty$ for $t\in[0,\infty)$. This assumption
is made plausible by our numerical results displayed in figure~\ref{Bild9}. 
Furthermore, we shall assume that the magnetic field itself is bounded, 
$|H(t)|\leq H_0$. It then follows from (\ref{gl:B:1}) that 
$|G(t)^2| \leq A(t) + H_0^2 M^2 t^2$. 
For long times, we may therefore expect a leading power-law behaviour which
we may write as $G(t) \sim t^{-w/2}$. We wish to estimate $w$ (the above
argument gives $w\geq -2$). 

\noindent {\bf Proposition:} {\it Let $G(t)$ be a solution of 
eq.~(\ref{gl:B:1}) and assume that there is a constant $M<\infty$ such 
that $|G(t)|\leq M$ for all times $t\in [0,\infty)$. Furthermore, the
oscillating field $H(t)$ is assumed to be bounded $|H(t)|\leq H_0$, piecewise
continuous and to have the Fourier expansion}
\BEQ \label{gl:B:2}
H(t) = H_0 \sum_{n=1}^{\infty} b_n \sin\left(\frac{2\pi n}{P} t\right)
\EEQ
{\it such that}
\BEQ \label{gl:B:3}
B := \sum_{n=1}^{\infty} \frac{1}{n} |b_n|
\EEQ
{\it is convergent. Finally, let $A(t)=f(t)=e^{-4dt}I_0(4t)^d$, where $I_0$
is a modified Bessel function. Then the exponent $w$ of the asymptotic form 
$G(t) \sim t^{-w/2}$ for $t\to\infty$ is for $P\ll 1$}
\BEQ
w = \frac{d}{2}
\EEQ

\noindent {\bf Proof:} 
First, for {\em any} magnetic field $H(t)$, we trivially have from 
eq.~(\ref{gl:B:1}) that $G(t)^2\geq A(t) = f(t) = (e^{-4t}I_0(4t))^d$ 
and therefore, as $t\to\infty$
\BEQ
G(t) \geq \frac{1}{(8\pi)^{d/4}} \cdot t^{-d/4}
\EEQ
Consequently, $w\leq d/2$. 

Second, we wish to find a sharp lower bound on $w$. This requires some
preparations, however. We begin by a discussion of the continuity of $G(t)$.
Let $\eps>0$ and consider
\BEA
\lefteqn{ G(t+\eps)^2 - G(t)^2 = 
\left( G(t+\eps) - G(t) \right) \left( G(t+\eps) + G(t) \right) }
\nonumber \\
&=& A(t+\eps) - A(t) + \int_{t}^{t+\eps} \!\D s\, H(s) G(s) 
\left(\int_{0}^{t+\eps}\!\D s\,H(s)G(s)+\int_{0}^{t} \!\D s\,H(s)G(s)\right)
\label{gl:B:5}
\EEA
Because of the first part and since $A(t)$ decreases monotonically with
$t$, we have $G(t+\eps)+G(t) \geq 2 \sqrt{A(t+\eps)\,}$ and obtain the estimate
\BEA
\left| G(t+\eps) - G(t) \right| &\leq& 
\eps \frac{|\dot{A}(t_A)|}{2\sqrt{A(t+\eps)\,}} + 
\int_{t}^{t+\eps} \!\D s\,\frac{\left|H(s)G(s)\right|}{2\sqrt{A(t+\eps)\,}}\;
\left(2\int_{0}^{t+\eps}\!\D s\, \left|H(s)G(s)\right| + {\rm O}(\eps)\right)
\nonumber \\
&\leq& \eps \left[ \frac{|\dot{A}(t)|}{2\sqrt{A(t)\,}} +  
\frac{H_0^2 M^2 t}{\sqrt{A(t)}} 
+ {\rm O}(\eps) \right]
\EEA
Here the mean value theorem was applied to $A(t)$ where $t_A$ is some 
intermediate value, $t_A\in [t, t+\eps]$. Taking the limit $\eps\to 0$, we
conclude that $G(t)$ is Lipschitz-continuous. Then we can apply the mean-value
theorem to eq.~(\ref{gl:B:5}). Because of the continuity of $G(t)$, the
limit
\BEQ
\lim_{\eps\to 0} \frac{G(t+\eps)-G(t)}{\eps} 
= \frac{\dot{A}(t)}{2G(t)} + H(t) \int_{0}^{t}\!\D s\, H(s) G(s)
\EEQ
exists for all times $t\in [0,\infty)$. Taking the derivative of (\ref{gl:B:1})
with respect to $t$, $G(t)$ satisfies the differential equation
\BEQ \label{gl:B:8}
\dot{G}(t) = \frac{\dot{A}(t)}{2G(t)} + H(t) \sqrt{ G(t)^2 - A(t)\:}
\EEQ
Therefore, if $H(t)$ is continuous, $\dot{G}(t)$ is also continuous. However,
if $H(t)$ has jumps, there may be jumps in $\dot{G}(t)$ as well. 

We shall take the average of eq.~(\ref{gl:B:8}) over the period interval  
$[t,t+P]$ of the external field. In order to prepare this, let $\vph(t)$ be 
a continuously differentiable function. Then
\BEA
\frac{1}{P} \int_{t}^{t+P} 
\!\!\D s\,\sin\left(\frac{2\pi n}{P}s\right)\vph(s)
&=& \frac{P}{2\pi n} \cos\left(\frac{2\pi n}{P}t\right) 
\frac{\vph(t)-\vph(t+P)}{P} + \frac{1}{2\pi n} 
\int_{t}^{t+P} \!\!\D s\,\cos\left(\frac{2\pi n}{P}s\right)\dot{\vph}(s)
\nonumber \\
&=& \frac{P}{2\pi n} \left[ 
-\cos\left(\frac{2\pi n}{P}t\right) \dot{\vph}(t_1)
+\cos\left(\frac{2\pi n}{P}t_2\right) \dot{\vph}(t_2) \right]
\EEA
where the mean-value theorems were applied and $t_1, t_2$ are intermediate
values from the interval $[t,t+P]$. For $P\to 0$, we expect $t_1, t_2\to \tau$. 
Since $-1\leq \cos x\leq 1$, we obtain the following bound for $P\ll 1$ 
\BEQ \label{gl:B:10}
\left| 
\frac{1}{P} \int_{t}^{t+P} \!\D s\,\sin\left(\frac{2\pi n}{P}s\right)\vph(s)
\right| \leq \frac{P}{\pi n} \left| \dot{\vph}(\tau)\right| + {\rm o}(P)
\EEQ

Before carrying out the average over eq.~(\ref{gl:B:8}), we consider the
approximation of $H(t)$ as given by eq.~(\ref{gl:B:2}) through a finite 
Fourier-sum $H_N(t) := \sum_{n=1}^{N} b_n \sin\left(\frac{2\pi n}{P} t\right)$. 
For every finite value of $N$, $H_N(t)$ is continuous and if we use $H_N(t)$ 
in eq.~(\ref{gl:B:8}), so is $\dot{G}(t)$. Furthermore, if $H(t)$ is
continuous, then $H_N(t)\rightrightarrows H(t)$ converges uniformly and the 
$N\to\infty$ limit and the integral may be interchanged, viz.
\BEQ \label{gl:B:11}
\lim_{N\to\infty} \int \!\D t\, H_N(t) \vph(t) = \int \!\D t\, H(t) \vph(t)
\EEQ
where $\vph(t)$ is some suitable function. If on the other
hand $H(t)$ is only piecewise continuous, we only have point-wise convergence
$H_N(t)\to H(t)$. In this case, the well-known Gibb's phenomenon occurs which
states that close to jump continuities the trigonometric approximations
$H_N(t)$ overshoot the limit $H(t)$ by about $9\%$ of the jump height,
see \cite{Cour68}. Because $H(t)$ is bounded, we certainly have the 
bound $|H_N(t)|\leq 3 H_0$ for all $N\in\mathbb{N}$ sufficiently large. 
Then the conditions of Lebesgue's theorem, 
see \cite{Cour68}, are satisfied and one arrives again at (\ref{gl:B:11}). 

Averaging eq.~(\ref{gl:B:8}) over a period interval, we obtain
\BEQ
\frac{1}{P} \int_{t}^{t+P} \!\D s\, \dot{G}(s) =
\frac{1}{P} \int_{t}^{t+P} \!\D s\, \frac{\dot{A}(s)}{2 G(s)} +
\frac{1}{P} \lim_{N\to\infty} \int_{t}^{t+P} \!\D s\, H_N(s) 
\sqrt{G(s)^2 - A(s)\,} 
\EEQ
We now consider $P\ll 1$. The left-hand side and 
the first term on the right-hand 
side are estimated by the mean-value theorem. For the second term, we use the 
inequality (\ref{gl:B:10}) term by term, taking
the $N\to\infty$ limit at the end. This gives, 
up to terms of order ${\rm o}(P)$
\BEA
\left| \dot{G}(\tau)\right| &\leq& 
\frac{1}{2} \left| \frac{\dot{A}(\tau)}{G(\tau)}\right| +
\left|\frac{\D}{\D\tau} \sqrt{G(\tau)^2-A(\tau)\,}\right| \frac{H_0 P}{\pi}
\sum_{n=1}^{\infty} \frac{|b_n|}{n}
\nonumber \\
&\leq& \frac{1}{2} \left| \frac{\dot{A}(\tau)}{G(\tau)}\right| +
\frac{H_0 P B}{\pi} \left| \frac{\dot{A}(\tau)}{2\sqrt{G(\tau)^2-A(\tau)\,}}
\right| + 
\frac{H_0 P B}{\pi} \frac{G(\tau)}{\sqrt{G(\tau)^2-A(\tau)\,}}
\left| \dot{G}(\tau)\right| 
\EEA
where the condition (\ref{gl:B:3}) was used. We finally arrive at
\BEQ \label{gl:B:14}
\left|\dot{G}(\tau)\right| \leq \left|\frac{\dot{A}(\tau)}{2G(\tau)}\right|
\Phi\left(\frac{H_0 P B}{\pi},\frac{A(\tau)}{G(t)^2}\right) \;\; ; \;\;
\mbox{\rm ~~where~~~}
\Phi(\alpha,x) := \left| 
\frac{1+\frac{\alpha}{\sqrt{1-x\,}}}{1-\frac{\alpha}{\sqrt{1-x\,}}}\right|
\EEQ

We now assume that the strict inequality $w<d/2$ holds. For large values of
$t$, we then have $A(t)/G(t)^2 \sim t^{w-d/2}\to 0$. On the other hand, the
function $\Phi(\alpha,x)$ has a p\^ole at 
$x_c=1-\alpha^2=1-(H_0 P B/\pi)^2$. There is a $t_0$ sufficiently large, 
such that, say, $A(t_0)/G(t_0)^2 \leq x_c/2$ and for $\tau\geq t_0$, we
have $\Phi(\alpha,A(\tau)/G(\tau)^2)\leq \Phi(\alpha,x_c/2)$, which is
finite. In addition, for uncorrelated initial conditions $A(t)=f(t)$, thus
$\dot{A}(t)=4de^{-4dt} I_0(4t)^{d-1}\left[ I_1(4dt)-I_0(4dt)\right]
\sim t^{-d/2-1}$. Inserting these asymptotic forms into eq.~(\ref{gl:B:14}),
we find $d/2\leq w$ in contradiction with the assumption $w<d/2$. 
\hfill q.e.d. 

The condition (\ref{gl:B:3}) is trivially satisfied for a sinusoidal field.
For a triangular field, one has $b_n\sim n^{-2}$ and for a rectangular field,
$b_n\sim n^{-1}$. In both cases $B$ is a finite constant and we have $w=d/2$,
in agreement with our numerical observation. 

The condition $P\ll 1$ is essential. For $P$ finite, numerical calculations
show that $G(t)$ goes towards a constant, modulated by a periodic function,
thus $w=0$. As $P$ becomes smaller, an intermediate regime appears, where
$G(t)\sim t^{-d/4}$, up to a modulation, before the regime mentioned above
is reached. This will be described in detail elsewhere. 

For any short-ranged initial condition, $A(t)\sim t^{-d/2}$ and our result
$w=d/2$ stays the same. On the other hand, for long-range initial correlations
of the form $C_{\rm ini}(\vec{r}) \sim |\vec{r}|^{-d-\alpha}$ with 
$\alpha<0$ \cite{Pico02}, we obtain in the same way $w=(d+\alpha)/2$. 

Finally, for a non-vanishing initial magnetization $S_0\neq 0$, we have
\BEQ
\dot{G}(t) = \frac{\dot{A}(t)}{2G(t)} + S_0 H(t) + H(t)\sqrt{G(t)^2 - A(t)\,}
\EEQ
Since $\int_{t}^{t+P} \!\D s\, H(s)=0$, the exponent $w=d/2$ is unchanged. 


\end{document}